\documentclass[preprint2,numberedappendix]{emulateapj-rtx4}
\usepackage{graphicx,natbib,color,bm,url,times}

\graphicspath{{./fig/}{./png/}}

\newcommand{\EQ}{\begin{equation}}
\newcommand{\EN}{\end{equation}}
\newcommand{\EQA}{\begin{eqnarray}}
\newcommand{\ENA}{\end{eqnarray}}
\newcommand{\eq}[1]{(\ref{#1})}
\newcommand{\EEq}[1]{Equation~(\ref{#1})}
\newcommand{\Eq}[1]{Equation~(\ref{#1})}
\newcommand{\Eqs}[2]{Equations~(\ref{#1}) and~(\ref{#2})}
\newcommand{\EEqs}[2]{Equations~(\ref{#1}) and~(\ref{#2})}

\newcommand{\Sec}[1]{Section~\ref{#1}}

\newcommand{\App}[1]{Appendix~\ref{#1}}
\newcommand{\Fig}[1]{Figure~\ref{#1}}
\newcommand{\FFig}[1]{Figure~\ref{#1}}
\newcommand{\Figs}[2]{Figures~\ref{#1} and \ref{#2}}
\newcommand{\Figss}[2]{Figures~\ref{#1}--\ref{#2}}
\newcommand{\Tab}[1]{Table~\ref{#1}}

\newcommand{\meanrho}{\overline{\rho}}

{}
{}
{}

{}
{}
{}
{}
{}
{}
{}
{}
{}
{}
{}
{}
{}
{}
{}
{}
{}
{}
{}
{}
\newcommand{\meanUU}{\overline{\mbox{\boldmath $U$}}{}}{}

{}
{}
{}

\newcommand{\meanU}{\overline{U}}

\newcommand{\meanS}{\overline{S}}
\newcommand{\meanT}{\overline{T}}
\newcommand{\meanP}{\overline{P}}

{}

{}
{}

\newcommand{\tbeta}{{\tilde{\beta}}}

\newcommand{\pphi}{\hat{\bm{\phi}}}

\newcommand{\HH}{\bm{H}}

\newcommand{\uu}{\mbox{\boldmath $u$} {}}
\newcommand{\UU}{\mbox{\boldmath $U$} {}}

\newcommand{\FF}{\mbox{\boldmath $F$} {}}

\newcommand{\grav}{\mbox{\boldmath $g$} {}}
\newcommand{\nab}{\mbox{\boldmath $\nabla$} {}}
\newcommand{\nabD}{\nabla_{\rm D}}
\newcommand{\nabad}{\nabla_{\rm ad}}
\newcommand{\nabkin}{\nabla_{\rm kin}}
\newcommand{\tnabad}{\widetilde\nabla_{\rm ad}}
\newcommand{\nabrad}{\nabla_{\rm rad}}

\newcommand{\ppsi}{\bm{\psi}}

\newcommand{\DD}{{\rm D} {}}

\newcommand{\dd}{{\rm d} {}}

\newcommand{\const}{{\rm const}  {}}

\def\la{\mathrel{\mathchoice {\vcenter{\offinterlineskip\halign{\hfil
$\displaystyle##$\hfil\cr<\cr\sim\cr}}}
{\vcenter{\offinterlineskip\halign{\hfil$\textstyle##$\hfil\cr<\cr\sim\cr}}}
{\vcenter{\offinterlineskip\halign{\hfil$\scriptstyle##$\hfil\cr<\cr\sim\cr}}}
{\vcenter{\offinterlineskip\halign{\hfil$\scriptscriptstyle##$\hfil\cr<\cr\sim\cr}}}}}
\def\ga{\mathrel{\mathchoice {\vcenter{\offinterlineskip\halign{\hfil
$\displaystyle##$\hfil\cr>\cr\sim\cr}}}
{\vcenter{\offinterlineskip\halign{\hfil$\textstyle##$\hfil\cr>\cr\sim\cr}}}
{\vcenter{\offinterlineskip\halign{\hfil$\scriptstyle##$\hfil\cr>\cr\sim\cr}}}
{\vcenter{\offinterlineskip\halign{\hfil$\scriptscriptstyle##$\hfil\cr>\cr\sim\cr}}}}}
\def\Ra{\mbox{\rm Ra}}
\def\Rat{\mbox{\rm Ra}_{\rm t}}
\def\Ratcrit{\mbox{\rm Ra}_{\rm t}^{\rm crit}}
\def\Ma{\mbox{\rm Ma}}

\def\Pra{\mbox{\rm Pr}}

\def\taured{\tau_{\rm red}}

\def\Teff{T_{\rm eff}}
\def\cp{c_{\rm p}}
\def\cv{c_{\rm v}}

\def\cs{c_{\rm s}}

\def\HP{H_{\!P}}
\def\cP{c_{P}}
\def\cV{c_{V}}

\def\kf{k_{\rm f}}
\def\kfz{k_{\rm f0}}

\def\sigmaSB{\sigma_{\rm SB}}

\def\urms{u_{\rm rms}}
\def\srms{s_{\rm rms}}

\def\nut{\nu_{\rm t}}

\def\chit{\chi_{\rm t}}

\def\onethird{{\textstyle{1\over3}}}

\newcommand{\K}{\,{\rm K}}
\newcommand{\g}{\,{\rm g}}
\newcommand{\s}{\,{\rm s}}

\newcommand{\cm}{\,{\rm cm}}

\newcommand{\m}{\,{\rm m}}
\newcommand{\km}{\,{\rm km}}

\newcommand{\Mm}{\,{\rm Mm}}

\newcommand{\dyn}{\,{\rm dyn}}

\newcommand{\Hminus}{{\rm H}^{-}}

\newcommand{\yjgr}[3]{ #1, {J.\ Geophys.\ Res.,} {#2}, #3}
\newcommand{\yjas}[3]{ #1, {J.\ Atmos.\ Sci.,} {#2}, #3}

\newcommand{\yapj}[3]{ #1, {ApJ,} {#2}, #3}

\newcommand{\yapjl}[3]{ #1, {ApJ,} {#2}, #3}

\newcommand{\yan}[3]{ #1, {Astron.\ Nachr.,} {#2}, #3}
\newcommand{\yzfa}[3]{ #1, {Z.\ Astrophys,} {#2}, #3}

\newcommand{\yana}[3]{ #1, {A\&A,} {#2}, #3}

\newcommand{\ypasj}[3]{ #1, {Publ.\ Astron.\ Soc.\ Japan,} {#2}, #3}
\newcommand{\yjfm}[3]{ #1, {J.\ Fluid Mech.,} {#2}, #3}

\newcommand{\ypf}[3]{ #1, {Phys.\ Fluids,} {#2}, #3}

\newcommand{\yanf}[3]{ #1, {Ann. Rev. Fluid Mech.,} {#2}, #3}

\newcommand{\yptrsa}[3]{ #1, {Phil. Trans. R. Soc. Lond. A,} {#2}, #3}
\newcommand{\ymn}[3]{ #1, {MNRAS,} {#2}, #3}

\newcommand{\ysci}[3]{ #1, {Science,} {#2}, #3}
\newcommand{\ysph}[3]{ #1, {Sol.\ Phys.,} {#2}, #3}

\newcommand{\ypra}[3]{ #1, {Phys.\ Rev.\ A,} {#2}, #3}

\newcommand{\ypnas}[3]{ #1, {Proc.\ Nat.\ Acad.\ Sci.,} {#2}, #3}

\newcommand{\yjcp}[3]{ #1, {J.\ Comput.\ Phys.,} {#2}, #3}
\newcommand{\yjour}[4]{ #1, {#2}, {#3}, #4}

\newcommand{\ybook}[3]{ #1, {#2} (#3)}
\newcommand{\yproc}[5]{ #1, in {#3}, ed.\ #4 (#5), #2}

\begin{document}
\preprint{NORDITA-2015-42}

\title{Stellar mixing length theory with entropy rain}

\author{Axel Brandenburg}
\affil{
Laboratory for Atmospheric and Space Physics, University of Colorado, Boulder, CO 80303, USA\\
JILA and Department of Astrophysical and Planetary Sciences, University of Colorado, Boulder, CO 80303, USA\\
Nordita, KTH Royal Institute of Technology and Stockholm University, Roslagstullsbacken 23, SE-10691 Stockholm, Sweden\\
Department of Astronomy, AlbaNova University Center, Stockholm University, SE-10691 Stockholm, Sweden
}

\submitted{\today,~ $ $Revision: 1.258 $ $}

\begin{abstract}
The effects of a non-gradient flux term originating from the motion
of convective elements with entropy perturbations of either sign are
investigated and incorporated into a modified version of stellar mixing
length theory (MLT).
Such a term, first studied by Deardorff in the meteorological context,
might represent the effects of cold intense downdrafts caused by the
rapid cooling in the granulation layer at the top of the convection zone
of late-type stars.
These intense downdrafts were first seen in the strongly stratified
simulations of Stein \& Nordlund in the late 1980s.
These downdrafts transport heat nonlocally, a phenomenon
referred to as {\em entropy rain}.
Moreover, the Deardorff term can cause upward enthalpy transport even in a
weakly Schwarzschild-stably stratified layer.
In that case, no giant cell convection would be excited.
This is interesting in view of recent observations, which could be explained
if the dominant flow structures were of small scale even at larger depths.
To study this possibility, three distinct flow structures are examined:
one in which convective structures have similar size and
mutual separation at all depths, one in which the separation
increases with depth, but their size is still unchanged, and one
in which both size and separation increase with depth,
which is the standard flow structure.
It is concluded that the third possibility with fewer and thicker
downdrafts in deeper layers remains the most plausible, but it may be
unable to explain the suspected absence of large-scale flows with speeds
and scales expected from MLT.
\end{abstract}

\keywords{
convection --- turbulence --- Sun: granulation
}

\section{Introduction}
\label{Intro}

Late-type stars such as our Sun have outer convection zones.
The observed solar granulation is a surface manifestation
of their existence.
Solar granulation and the first few megameters (Mm) of the
convection zone have been modeled successfully
using mixing length theory (MLT) and numerical simulations with realistic
physics included \citep{SN89,SN98,VSSCE05,Gudiksen,Freytag12}.
As a function of depth, the simulations reproduce some essential
features predicted by MLT, in particular
the depth dependence of the turbulent
root mean square (rms) velocity,
$\urms\approx(F_{\rm enth}/\rho)^{1/3}$, where $F_{\rm enth}$
is the enthalpy flux and $\rho$ is the density.
Simulations also seem to confirm an important {\em hypothesis}
of MLT regarding the gradual increase of the typical convective time
and length scales with depth.
Given such agreements, there was never any reason to question
our basic understanding of convection.

In recent years, local helioseismology has allowed us to determine
subsurface flow velocities \citep{Duvall97,GB05}.
Helioseismic observations by \cite{HBBG10,Hanasoge} using the
deep-focusing time-distance technique have not, however, detected
large-scale convection velocities at the expected levels \citep{MFRT12};
see also \cite{GB12} for a comparison with both global simulations
and radiation hydrodynamics simulations with realistic radiation and
ionization physics included.
\cite{Greer} have suggested that the approach of \cite{Hanasoge}
may remove too much signal over the time span of the measurements.
Using instead ring diagram analysis \citep{GB05} with appropriately
assembled averaging kernels to focus on the deeper layers,
\cite{Greer} instead find values of the turbulent rms velocities
that are consistent with conventional wisdom.
Moreover, they find that at large length scales corresponding to
spherical harmonic degrees of 30 or below (corresponding to
scales of about $140\Mm$ and larger), the rms velocity actually
{\em increases} with depth.
This in itself is remarkable, because velocity perturbations from
deeper layers were expected to be transmitted to the surface almost
unimpededly \citep{Sti81}, unlike perturbations in the heat flux,
which are efficiently being screened by the convection \citep{Spr77}.
Thus, an increase in horizontal velocity power with depth
is unexpected.
However, \cite{vanBall86} pointed out that for the scale and depth
of giant cells, the screening might be large enough to allow giant cell
convection of $100\m\s^{-1}$ to result in only $10\m\s^{-1}$ at the
surface, which would be compatible with observations \citep{Hathaway}.
The presence of the near-surface shear layer of the Sun \citep{Schou98},
might enhance the screening further.

Given that the deeply focused kernels used by \cite{Greer} can still
have 5--10\% sensitivity to Doppler shifts arising from flows in the
upper layers, such near-surface flows could still leave an imprint on
the signal.
This effect would be exaggerated further if the near-surface flows
were stronger compared to those in deeper layers, as was
theoretically expected.
On the other hand, if convection in the surface layers is of smaller
scale, the signal from those layers will to a large extent be averaged
out despite its larger amplitude.
It would obviously be important to examine this more thoroughly
by applying the kernels of \cite{Greer} to realistic simulations.
Unfortunately, this has not yet been attempted.
Conversely, the deep-focusing technique of \cite{Hanasoge} could be
extended to allow for imaging of the deeper flow structures and
thereby a direct comparison of individual turbulent eddies with those
detected by \cite{Greer}.

It should be mentioned that small flow speeds of giant cell convection
have been found by correlation tracking of supergranule proper
motions \citep{Hathaway}.
The typical velocities are of the order of $20\m\s^{-1}$ at spherical
degrees of around 10.
Those speeds are still an order of magnitude above the helioseismic upper
limits of \cite{Hanasoge}, but \cite{HS14} argue that these surface
measurements translate to lower flow speeds in the deeper and denser
layers when considering mass conservation.
This argument may be too naive and would obviously be in conflict with
the results of \cite{Greer}, which show an {\em increase} with depth at
low spherical harmonic degrees.
On the other hand, \cite{BBB15} have argued that the flows reported by
\cite{Hathaway} may be non-convective in nature and, in fact, magnetically
driven and perhaps related to the torsional oscillations.

Quantitatively, the horizontal flow speed at different scales is
characterized by the kinetic energy spectrum, $E(k)$, where $k$ is the
wavenumber or inverse length scale.
\cite{SBN07} have shown that the spectra of surface Dopplergrams
and correlation tracking collapse with those of Stein \& Nordlund
onto a single graph such that the spectral velocity $[kE(k)]^{1/2}$
is proportional to $k$, i.e., $E(k)\propto k$; see also \cite{NSA09}.
This is a remarkable agreement between simulations and observations.
While the origin of such a spectrum is theoretically not
understood, it should be emphasized that these statements concern
the horizontal velocity at the surface and do not address the
controversy regarding the spectrum at larger depths, where the flow speeds
at large length scales may still be either larger \citep[e.g.][]{Greer}
or smaller \citep[e.g.][]{FH16} than those at the surface.
If most of the kinetic energy were to reside on small scales throughout
the convection zone even at a depth of several tens of megameters and
beneath, $E(k)$ is expected to decrease toward smaller wavenumbers $k$
either like white noise $\propto k^2$ or maybe even with a Batchelor
spectrum $\propto k^4$ \citep{Dav04}.
However, even in simulations of forced isothermally stratified turbulence,
in which there is no larger-scale driving from thermal buoyancy,
there is a shallower spectrum, $E(k)\propto k^{3/2}$ \citep{Los13}.
By contrast, if one imagines the flow to be anelastic so that the mass
flux $\rho\uu=\nab\times\ppsi$ can be written as a vectorial stream
function $\ppsi$, and if $\ppsi$ is given by white noise, one should
expect a $k^4$ Batchelor spectrum.
This is qualitatively similar to the results of granule tracking,
which reveal an intermediate scaling proportional to $k^3$
\citep{Rieutord08,Roudier12}.
While the steeper $k^4$ spectrum for $k<\kf$ might leave some hope that
the results of \cite{Hanasoge} could be reconciled with these theoretical
and observational constraints, this would be virtually impossible for
the linear energy spectrum, $E(k)\propto k$.

The issue has recently been examined by \cite{Lord}, who have shown that
simulation results with the MURaM code \citep{VSSCE05} can be reproduced
with a model that is composed of a continuous hierarchy of layers, each
with its own driving scale at a scale of four local density scale heights.
At wavenumbers below that driving scale, the spectral power falls off
in a certain way.
The horizontal surface spectrum is a superposition of these
contributions, each of which is assumed to decay with height, contributing
therefore progressively less with depth.
However, both their model and the MURaM results \citep{Lord14} show
an order of magnitude more power (factor $2.5$ in rms velocity)
than the Sun at small wavenumbers.
To understand this, they show that the observations can be reproduced
if below a depth of $10\Mm$ the convective energy flux is augmented
by an artificially added flux term, which thus significantly reduces the
rms velocity of the resolved flow field.
They speculate that the flow--temperature correlation entering the
enthalpy flux may be larger in the Sun, possibly being caused by a
magnetic field that may maintain flow correlations and boost the
transport of convective flux by smaller scales.
This has been partially confirmed by \cite{Hotta}, who have discussed
the possible role of small-scale magnetic fields in suppressing the
formation of large-scale flows.
Furthermore, \cite{FH16} have found that with increasing Rayleigh numbers,
there is more kinetic energy at small scales and less at large scales
such that the total kinetic energy is unchanged by this rearrangement
of energy.

Given the importance of this subject, 
it is worthwhile reviewing possible shortcomings in our theoretical
understanding and numerical modeling of stellar convection.
Both global and local surface simulations of solar-type convection would
become numerically unstable with just the physical values of viscosity
and radiative diffusivity.
Therefore, the radiative flux (in the optically thick layers) is modified
in one of two possible ways.
(i) The contribution from temperature fluctuations is greatly enhanced, i.e.,
\EQ
\FF_{\rm rad}=-K\nab T \quad \rightarrow \quad
-K\nab\meanT-K_{\rm SGS}\nab(T-\meanT),
\label{FradSplit}
\EN
where $T$ and $\meanT$ are the actual and horizontally averaged
temperatures, respectively, $K$ is the radiative conductivity,
and $K_{\rm SGS}$ is a subgrid scale (SGS) conductivity.
The latter is enhanced by many orders of magnitude relative to $K$.
Furthermore, numerical diffusion operators often do not translate in any
obvious way to the physical operators.
(ii) Alternatively, in {\em direct} numerical simulations (DNS), one uses
physical viscosity and diffusivity operators, i.e., the replacement
in \Eq{FradSplit} is not invoked, but the coefficients are enhanced
($K\to K_{\rm enh}$) and exceed the physical
ones by many orders of magnitude.
Both approaches are problematic.

In many global DNS with enhanced coefficients \citep[e.g.,][]{KMCWB13},
the lower boundary is closed, so at the bottom of the domain all the
energy is carried by radiation alone.
By choosing an enhanced radiative diffusivity, the radiative flux is
increased by a corresponding amount, and therefore also the total flux.
The luminosity in those simulations can exceed the solar value
by several orders of magnitude.
This has a series of consequences.
Most notably, the convective velocities are too high in the upper
parts where most of the flux is carried by convection \citep{KMCWB13}.
There are two ways to avoid this problem.
One is to use simulations with a polytropic hydrostatic reference
solution, whose polytropic index is close enough to the adiabatic one
so that the convective flux is everywhere a small fraction of the
radiative one \citep{BCNS05}, which reduces the convective velocities.
Alternatively, one applies the enhanced radiative conductivity only
to the temperature fluctuations so as not to disturb the very small
radiative energy flux, $-K\nab T$, compared with $-K_{\rm enh}\nab T$,
which it would have been in the DNS approach without subtracting $\meanT$.
The temperature smoothing implied by invoking alternative (i)
is also necessary in the simulations with realistic opacities,
because the P\'eclet number (which is similar to the Reynolds number)
based on the physical value of $K$ would reach values above
$10^{10}$, which cannot be handled by a DNS \citep{BB14}.
In the global simulations, it is common to use the specific entropy gradient
instead of $\nab(T-\meanT)$ \citep[e.g.][]{KMCWB13}.
This is equivalent if $\meanT$ is close to the adiabatic value.
However, the diffusion coefficient in the SGS term can easily
be five orders of magnitude larger than the physical one acting on the
mean stratification.
This has the consequence of suppressing small-scale turbulent flows
and entropy fluctuations, which may prematurely damp the
low-entropy fluid parcels that originate within the strong cooling
layer at the surface and which will be discussed as {\em entropy rain}
in the bulk of this paper.
If such low-entropy contributions are poorly captured by the simulation,
this could be compensated for by a sufficiently strong contribution
proportional to the superadiabatic gradient, which in turn would give
rise to the excitation of flows on much larger scales than what would
be compatible with the observations discussed by \cite{Lord}.
The sensitivity of large-scale flow excitation to small changes
in the superadiabatic gradient was also discussed by \cite{CR16},
who studied models with different types of surface driving.
It is further supported by the work of \cite{FH16}, who found that there
is more kinetic energy at small scales and less at large scales as the
Rayleigh number increases.

There is yet another problem that concerns the global simulations
in which a predetermined profile $K(z)$ is used instead of
calculating its local value with a physical opacity.
Such an approach has been used routinely in studies of compressible
convection, especially when stably and unstably stratified layers
are combined \citep{HTM86,BJNRST96}.
The simulations of \cite{KMCWB13} adopted a profile that yields
a negative (unstable) radial entropy gradient through most of
the layer (see the inset of their Fig.~3).
However, as will be pointed out below, it is only a tiny surface
layer in which the non-convective model is unstable.
The rest of the model is a priori stable to convection, but it becomes
marginally stable (or perhaps slightly unstable) as a consequence of
the resulting turbulence leading to bulk mixing across the deeper layers.
This is quite contrary to the models with a predefined $K(z)$ profile,
which would be unstable by construction over the entire depth of the
convection zone.

We should emphasize that the non-convective solution is mainly of
academic interest.
It is used to compute, for example, the Rayleigh number, a measure of
the degree of instability.
However, there is no doubt that the actual stratification of the
Sun is close to marginal stability down to a fractional radius of 0.71
before turning decidedly stable,
as confirmed by helioseismology \citep{CDGT91,Basu}.
Nevertheless, the concept of convection being driven by surface cooling
rather than heating from below has been promoted in a number of papers
by \cite{SN89,SN98} and elaborated upon by \cite{Spr97}, who introduced
the idea that flows in the deeper parts are being driven nonlocally
through the entropy rain from the surface.
The question is to what extent this affects our understanding of the
speed and especially the typical scales of convective motions, how MLT
models would need to be modified, and whether this might have any bearing
on the interpretation of the observed flow amplitudes at large scales.

To include the nonlocal effects described by \cite{Spr97}, we
must look for a contribution to the enthalpy flux that is not related
to the local entropy gradient.\footnote{As will be discussed in
\Sec{UnstablyStratified}, \cite{Spr97} argues that the low-entropy
material from the top always leads to a negative mean entropy gradient.
However, here we argue that the mean entropy gradient is only one
contribution to a mean-field (here one-dimensional)
parameterization of the enthalpy flux, and that
there is another one that is not locally connected with it.}
Such a term has been identified in the meteorological context by \cite{Dea66},
who describes it as a counter-gradient flux.
In \cite{Dea72}, he derives an expression for this flux, which depends
on the local temperature or entropy fluctuations which, in his case,
come from measurements.
In the present case, we assume that such fluctuations have their origin in
what \cite{Spr97} refers to as {\it threads}, which are thin downdrafts on
an almost perfectly isentropic passive fluid upflow between the threads.
Spruit already emphasized back then (p.\ 406) the dynamical importance of the
``forest of narrow cool threads that are produced at the solar surface.''
He further suggested (p.\ 406) that their absence in simulations with a fixed top
boundary at some depth below the actual surface might be ``the reason why
they would produce large-scale flows with amplitudes that are about
two orders of magnitude larger than observed.''

In the scenario described by \cite{Spr97}, convection is driven solely
within the surface layers.
This is superficially reminiscent of convective overshoot, as modeled
in some of the aforementioned papers \citep{HTM86,BJNRST96},
where convective flows are driven into stable layers.
The extent of overshoot depends not only on the
stiffness of the stable gradient that the convective plumes are
flowing into \citep{HTMZ94}, but also on the flow speed.
The depth of such a layer can therefore not
be determined a priori from purely hydrostatic stability considerations.
The lower boundary might then not be sharp.
As explained above, in a hydrostatic non-convecting reference model,
only a very thin surface layer is a priori unstable to convection.
The rest is made unstable purely by bulk mixing.
If entropy rain convection is a nonlocal phenomenon, the extent of
convection should depend on surface properties and cannot be predicted
from the local entropy gradient, similarly to convective overshoot.
One might then not be able to understand the relatively sharp demarcation
at the bottom of the convection zone, as found in global helioseismology.
Of course, once convection has become fully developed, the low
entropy elements descend into buoyantly neutral layers, which is quite
different from the usual overshoot.
Furthermore, the usual overshoot layer is characterized by negative
buoyancy and therefore a downward enthalpy flux \citep{HTM86}.
An important purpose of the present paper is to produce
a quantitative model and to demonstrate that,
with a hypothetical nonlocal contribution to the flux, it is possible
to obtain models that still have a sharply defined lower boundary.

A quantitative model of convection with entropy rain, even with the
somewhat hypothetical Deardorff term included, would also be useful to
illustrate the qualitative nature of the resulting stratification,
and to show whether it is weakly super or subadiabatic.
Indeed, as already argued by \cite{Spr97}, if the local
mass fraction of entropy deficient material, descending from the
cooling surface, decreases with depth and if the stratification
outside the entropy rain were exactly isentropic, the resulting
horizontally averaged entropy would {\em increase} with depth.
This suggests that the entropy rain itself can make
an otherwise vanishing radial entropy gradient negative
and therefore Schwarzschild unstable, as seen in surface simulations.
This raises two important questions.
First, to what extent is such a stratification affected by radiative
heating---especially toward the bottom of the convection zone,
where it would tend to produce a positive (stable) mean entropy
gradient in the upflows. Second, would such a negative (unstable)
mean entropy gradient always lead to giant cell convection, as has been
seen in global convection simulations \citep{Miesch08}, or could the
stratification still be stable to the excitation of large-scale flows?

We postpone addressing the two aforementioned issues connected with the
qualitative reasoning of \cite{Spr97} to the end of the paper
(\Sec{UnstablyStratified}), and begin by highlighting the less
commonly known fact that in the non-convecting reference model, only the
layer in the top $1\Mm$ is convectively unstable (\Sec{HighlyUnstable}).
We then explain the nature of the Deardorff flux (\Sec{Deardorff}),
present a correspondingly modified mixing length model (\Sec{Modified})
and give some illustrative numerical solutions (\Sec{NumSol}).
We discuss alternative explanations for the lack of giant cell convection in
\Sec{UnstablyStratified} and present our conclusions in \Sec{Conclusions}.

\section{A highly unstable surface layer}
\label{HighlyUnstable}

The main argument for the existence of a highly unstable surface layer
comes from the consideration of the associated non-convective
reference solution, where the flux is forced to be
transported by radiation only, i.e.\ $F=F_{\rm rad}$.
This is something that is not normally considered in stellar physics,
because we know that such a solution would be unstable to convection
and would therefore never be realized.
However, as mentioned in the introduction, this solution is of certain
academic interest.
We postpone the discussion of an explicit numerical solution to \Sec{NumSol}
and present in this section the basic argument only at a qualitative
level, making reference to earlier numerical calculations by \cite{BB14}
for a simple model.
They considered an opacity law of the form
\EQ
\kappa=\kappa_0 (\rho/\rho_0)^a (T/T_0)^b,
\label{Kramers}
\EN
where $a$ and $b$ are adjustable parameters, $\rho_0$ and $T_0$ are
reference values for density and temperature, respectively, and $\kappa_0$
gives the overall magnitude of the opacity.
The essential point to note here is that the exponents $a$ and $b$ determine
the gradient of specific entropy in a purely non-convecting reference model.
In thermodynamic equilibrium, the radiative flux must be constant, i.e.,
\EQ
F_{\rm rad}=-K\,\dd T/\dd z=\const,
\label{FradEqn}
\EN
where $K=16\sigmaSB T^3/(3\kappa\rho)$ is the radiative conductivity
with $\sigmaSB$ being the Stefan--Boltzmann constant, and $z$ is the
vertical coordinate in a Cartesian coordinate system.
For the simple opacity law \eq{Kramers}, but with
\EQ
b<4+a,
\label{bpoly}
\EN
the optically thick regime is characterized by a constant
temperature gradient and therefore $K=\const$ (see \App{Kconst}).
We have then a polytropic stratification with $\rho\propto T^n$, where
\EQ
n=(3-b)/(1+a)
\label{n_Def}
\EN
is the polytropic index.
It is larger than $-1$ when \Eq{bpoly} is obeyed, i.e., when the pressure
decreases with height, as is required for a physically meaningful solution.
For a ratio of specific heats of $\gamma=5/3$, the value of $n$ for marginal
Schwarzschild stability is $n_{\rm crit}=1/(\gamma-1)=3/2$.
In most of the solar convection zone the dominant opacity is the
bound-free absorption owing to the absorption of light during ionization
of a bound electron, which is well described by the Kramers-type opacity law
with $a=1$ and $b=-7/2$, so $n=3.25$, which corresponds to a
Schwarzschild-stable solution.
Only near the surface, at temperatures typically below $15,000\K$, the
dominant opacity is the $\Hminus$ opacity.
It can no longer be approximated by a simple power law of the
type given by \Eq{Kramers}.
However, in limited density and temperature ranges, certain values
of $a$ and $b$ can tentatively be specified, e.g., $a=0.5$ and $b=7...18$.
Clearly, for all these values the constraint \eq{bpoly} is violated, so the
hydrostatic stratification is no longer polytropic,
but solutions can still be constructed numerically \citep{BB14} and they
demonstrate, not surprisingly, that the stratification is highly unstable.
What is more surprising is the fact that even with a combined opacity law
of the form
\EQ
\kappa^{-1}=\kappa^{-1}_{\rm Kr}+\kappa^{-1}_{\Hminus},
\label{doubleKramers}
\EN
where $\kappa_{\rm Kr}$ and $\kappa_{\Hminus}$ are given by
\Eq{Kramers} with suitable exponents $a$ and $b$,
the solutions of the non-convective reference state is unstable only
over a depth of approximately $1\Mm$.
We return to this at the end of \Sec{NumSol},
where we present numerical solutions.

Of course, as stated earlier, the non-convective reference state
is only of academic interest.
Already with standard MLT \citep{Bie32,Vit53}, which allows for a
non-vanishing enthalpy flux, one finds a vastly extended convection zone
with a depth of the order of $100\Mm$ \citep{Bie38}.
However, the question now is how this can be affected by the presence of
the Deardorff flux.
This will be the subject of the rest of this paper.
If the Deardorff flux were to become dominant and the
stratification subadiabatic, it would become locally stable to the
onset of convection.
One might further speculate that the typical scale would
no longer be controlled by the local pressure scale height, but
it might be imprinted from the downdraft pattern just beneath the surface
and be therefore comparable to the granulation scale or at least
the supergranulation scale \citep{CR16}.
This can have other important consequences that will also be addressed
in this paper.
It should be pointed out, however, that the surface simulations
have so far not produced evidence for subadiabatic stratification.
On the other hand, the models presented below predict subadiabatic
stratification only a certain distance below the surface,
depending on ill-known input parameters.

\section{The Deardorff flux}
\label{Deardorff}

\subsection{Derivation}
\label{Derivation}

In the meteorological context, counter-gradient heat flux terms have been
noticed for a long time \citep{Ert42,PS47,Dea66}.
They appear naturally when calculating the enthalpy flux $F_{\rm enth}$
using the $\tau$ approximation in its minimalistic form \citep[e.g.][]{BF03}.
In this approach, one computes the time derivative of $F_{\rm enth}$.
In the absence of ionization effects, $F_{\rm enth}$ can be written as
\EQ
F_{\rm enth}=\overline{\rho u_z \cP T},
\label{Fconv_orig}
\EN
where the overbar denotes horizontal averaging.
In standard MLT, the enthalpy flux is usually referred to
as the convective flux and the kinetic energy flux vanishes
because of the assumed perfect symmetry between up- and downflows.
In deeper layers especially, however, this is not justified
\citep{SNGBS09}, so this restriction will later be relaxed.

In the case of strongly stratified layers, it is convenient to use
pressure $P$ and specific entropy $S$ as thermodynamic variables, because
we later want to ignore pressure fluctuations on the grounds that pressure
disturbances are quickly equilibrated by sound waves.
Furthermore, we will restrict ourselves to second order correlations
in \Eq{Fconv_orig} and thus to fluctuations only in the correlation of
specific entropy and velocity.
Up to some reference value, we have
\EQ
S/\cP=\ln T-\nabad\ln P,
\label{Seqn}
\EN
where $\nabad=1-1/\gamma$, and $\gamma=\cP/\cV$ is the ratio of specific
heats at constant pressure and constant volume, respectively,
all constants for a perfect gas with a fixed degree of ionization.
Ignoring pressure variations and linearizing $\delta\ln T=\delta T/T$,
we can replace $\cP\delta T$ by $T\delta S$.
In the following, we denote fluctuations by lower case characters,
i.e., $S=\meanS+s$, where $s\equiv\delta S$ are used interchangeably.
As argued above, other fluctuations are omitted.
We also ignore mean flows ($\meanUU=\bm{0}$), so there are only velocity
fluctuations ($\UU=\uu$).
Focusing thus on the dominant contribution proportional to
$\overline{u_z s}$, we have
\EQ
F_{\rm enth}=\meanrho\, \meanT\,\overline{u_z s}.
\label{Fenth}
\EN
Next, we write the time derivative of $F_{\rm enth}$ as
\EQ
\partial F_{\rm enth}/\partial t=\meanrho\, \meanT
\left(\,\overline{u_z \dot{s}}+\overline{\dot{u}_z s}\,\right),
\label{dFconv}
\EN
where dots denote partial time derivatives and changes
of the background state have been neglected.
Using the governing equations for $u_i$ and $s$, we have
(see \App{deriv10and11})
\EQA
&&\dot{s}=-u_j\nabla_j\meanS-s/\tau_{\rm cool}...,
\label{sdot}
\\
&&\dot{u}_i=-g_i s/\cP+...,
\label{udot}
\ENA
where $g_i$ is the $i$th component of the gravitational acceleration
in Cartesian coordinates, namely $\grav=(0,0,-g)$, the ellipses
refer to terms that are nonlinear in the fluctuations,
\EQ
\tau_{\rm cool}^{-1}=\iota c_\gamma\kf
\;\;\mbox{with}\;\; \iota=\ell\kf/(3+\ell^2\kf^2)
\label{taucool}
\EN
is the inverse heating and cooling time owing to radiation
\citep{US66,Edw90}, $c_\gamma=16\sigmaSB \meanT^3/\meanrho \cP$
is the photon diffusion speed \citep{BB14},
$\ell=1/\kappa\rho$ is the photon mean-free path, and $\kf$ is the typical
wavenumber of the fluctuations, which may be associated with the
inverse mixing length used routinely in MLT.
We mention at this point that the nonlocal nature of the entropy rain
must lead to yet another contribution in \Eq{sdot}.
We expect this to be the result of the nonlinear term
(indicated by the ellipses),
of which a part later gives rise to the $\nabD$ term.
This will be motivated further at the end of this section
and in \Sec{DepthDependence}, where it will be included
in the final expression for $\nabD$.

Using \Eqs{sdot}{udot} in \Eq{dFconv}, we have
\EQ
{\partial F_i^{\rm enth}\over\partial t}\!=\!
\meanrho\, \meanT \left(-\overline{u_i u_j}\,\nabla_j\meanS
-g_i \overline{s^2}/\cP \right)-{F_i^{\rm enth}\over\tau_{\rm cool}}
+{\cal T}_i,
\label{dotFconv}
\EN
where the $\overline{s^2}$ term is the Deardorff flux and
the ${\cal T}_i$ refer to triple correlations that will be approximated
by the quadratic correlation $F_{\rm enth}$ as
\EQ
{\cal T}_i=-F_i^{\rm enth}/\tau,
\label{MTAclosure}
\EN
where $\tau$ is a relaxation time due to turbulence \citep{BF03},
which will later be identified with the turnover time.
This procedure, in which correlations with pressure fluctuations
are also neglected, is called the minimal $\tau$ approximation.
Important aspects of the closure assumption \eq{MTAclosure} have been
verified numerically for passive scalar transport \citep{BKM04}.
Among other things, they found that the time derivative on the
left-hand side of \Eq{dotFconv} restores causality by turning the
otherwise parabolic heat equation into a hyperbolic wave equation.
Here, however, we are interested in slow variations such that
the time derivative of $F_{\rm enth}$ can be neglected.
(Note, however, that this does not imply that the cooling term
will be neglected.)

Since the relaxation term in \Eq{MTAclosure} is similar to the cooling
term in \Eq{dotFconv}, we can combine the two by introducing 
the reduced relaxation time $\taured$ defined through
\EQ
\taured^{-1}=\tau_{\rm cool}^{-1}+\tau^{-1}.
\label{tau}
\EN
We can then solve for $F_{\rm enth}$, which appears on the
right-hand sides (rhs) of \Eqs{dotFconv}{MTAclosure}, and find
$\FF_{\rm enth}=\FF_{\rm G}+\FF_{\rm D}$, where
\EQA
\label{FG}
\FF_{\rm G}
&=&
-\onethird\taured\urms^2
\,\meanrho\,\meanT\,\nab\meanS,
\\
\label{FD}
\FF_{\rm D}
&=&
-\taured\overline{s^2}\,\grav\,\meanrho\,\meanT/\cP
\ENA
are the ordinary gradient and the new Deardorff fluxes,
respectively, and anisotropies have been ignored for the benefit of
simpler notation.
Thus, we write $\overline{u_i u_j}\approx\onethird\delta_{ij}\urms^2$,
where $\urms$ is the rms velocity of the turbulence.
\EEqs{FG}{FD} are written in vectorial forms to highlight the directions
of the fluxes: $\FF_{\rm G}$ is counter-gradient and $\FF_{\rm D}$
is countergravity.
In \Eq{FG}, the term $\onethird\taured\urms^2\equiv\chit$ is
the turbulent thermal diffusivity.

\begin{figure}[t!]\begin{center}
\includegraphics[width=\columnwidth]{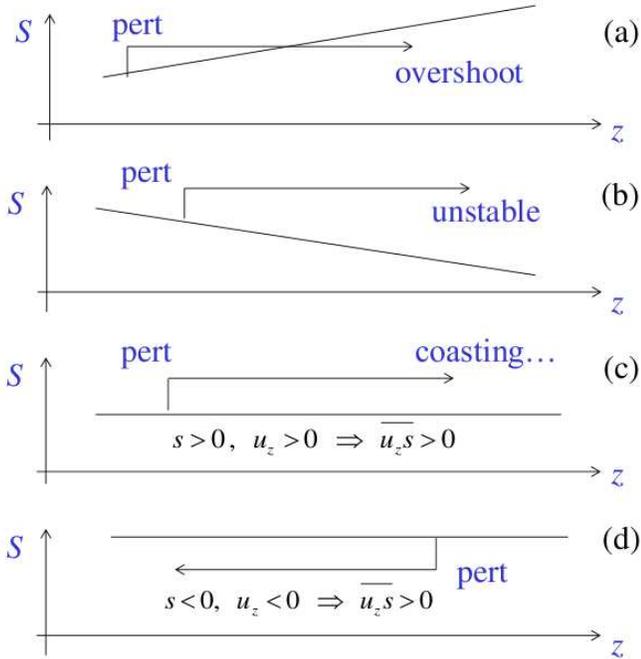}
\end{center}\caption[]{
Sketch illustrating overshoot in a stably stratified layer (a),
growth of perturbations in an unstably stratified layer (b),
buoyant rise of a blob with positive entropy perturbation (c), and
descent of a blob with negative entropy perturbation and hence
negative buoyancy (d).
In cases (b)--(d), the turbulent flux is upward (increasing $z$).
Case (d) is relevant to entropy rain.
}\label{sketch}\end{figure}

In view of astrophysical applications, we replace the specific entropy
gradient by the commonly defined superadiabatic gradient, i.e.,
\EQ
-\dd(\meanS/\cP)/\dd z=(\nabla-\nabad)/\HP,
\EN
where $\nabla=\dd\ln \meanT/\dd\ln \meanP$ is the double-logarithmic
temperature gradient and $\HP=-(\dd\ln\meanP/\dd z)^{-1}$ is the
pressure scale height.
Thus, we arrive at
\EQ
F_{\rm enth}=\onethird\meanrho \cP \meanT \, (\taured\urms^2/\HP)
\left(\nabla-\nabad+\nabD\right),
\label{Fconv_urms2}
\EN
where $\nabD$ is a new contribution to standard MLT, which
results from $F_{\rm D}$.
Using the terms on the rhs of \Eq{FD}, we can write it explicitly as
\EQ
\nabD=(3/\gamma)\, (\overline{s^2}/\cP^2)\; \Ma^{-2},
\label{nabD}
\EN
where $\Ma=\urms/\cs$ is the Mach number of the turbulence and $\cs$ is
the sound speed with $\cs^2=\gamma g\HP$.
Note that this Mach number dependence arises in order to cancel the
corresponding $\urms^2$ factor in the definition \eq{Fconv_urms2}.
The Deardorff term is a contribution to the flux that is always directed
outward and results from the transport of fluid elements with entropy
fluctuations of either sign, as is illustrated in \Fig{sketch}.

\subsection{Physical interpretation}
\label{Interpretation}

We recall that in a stably (unstably) stratified layer, shown in
\Fig{sketch}(a) and (b), and under the assumption of pressure equilibrium,
the entropy perturbation of a blob with respect to its current surroundings
decreases (increases) after a small ascent.
By contrast, in the marginally or nearly marginally stratified cases,
the perturbation remains constant if one ignores mixing
with the surroundings; see panels (c) and (d).
Therefore, a positive entropy
perturbation ($s>0$) always corresponds to a positive temperature
perturbation and a negative density perturbation.
Thus, the fluid parcel is buoyant and moves upward ($u_z>0$), so $u_zs>0$,
giving a positive contribution to $F_{\rm enth}$.
Likewise, for $s<0$, the parcel is cooler and heavier and sinks ($u_z<0$),
and again, $u_zs>0$, giving a positive contribution to $F_{\rm enth}$.

In laboratory convection, strong positive entropy perturbations can
be driven at the lower boundary, and strong negative ones at the top
boundary.
In the Sun, however, only the strong radiative cooling in the
photosphere provides a significant source of entropy perturbations.
This became clear with the emergence of the realistic solar convection
simulations of \cite{SN89}.
These simulations showed for the first time the strong vertical asymmetry
of solar convection resulting from the physics of ionization and strong
cooling via the $\Hminus$ opacity.
This led to the notion of {\em entropy rain} and Spruit's description
of solar convection as a nonlocal phenomenon that is driven solely by
surface cooling.
Our considerations suggest that the resulting
(one-dimensional) mean-field energy flux is
parameterized not just in terms of the local superadiabatic gradient.

In analogy with laboratory convection \citep{HCL87,Castaing}, Spruit refers
to the downdrafts as threads.
Interestingly, the laboratory experiments show that these threads persist
even at large Rayleigh numbers.
For such threads to persist, one could imagine them to be stabilized
by their intrinsic vorticity, akin to a Hill vortex \citep{Hil94}.
These are vortex rings that can stay concentrated over large distances.
Vortex rings have also been reported by \cite{SNGBS09}, but many of
those are associated with mushroom shapes, indicating that they widen
and soon break up and mix with their surroundings.
Numerical studies (see \App{HillVortices}) suggest that Hill vortices
also survive in a highly stratified isothermal layer, and that their
persistence increases significantly with resolution and decreasing
viscosity.
On the other hand, the presence of background turbulence provides an
effective turbulent viscosity, which contributes to a premature
break up of small-scale vortices.
However, those studies were done with an isothermal equation of state
and thus ignore the effects of continued driving by a persistent negative
entropy perturbation between the vortex and its surroundings.

Studies of the Deardorff flux in the meteorological context suggest that
the flux-carrying plumes are associated with so-called coherent structures
\citep{DeRoode04} and those may not simply be the vortex-like structures
envisaged above.
It is clear, however, that the structures seen in the Earth's
atmosphere are driven by the boundary layer on the ground, and sometimes
also by the upper inversion layer \citep{DeRoode04}.
\cite{Ple07} discusses the Deardorff flux in connection with other
parameterizations of nonlocal fluxes such as the
transilient matrix approach \citep{Stu84,Stu93}.
The close connection between counter-gradient fluxes and nonlocal
transport has been elaborated upon by \cite{vDop01} and \cite{Buske07}.
Thus, while the approach presented in \Sec{Derivation} may capture the
physical phenomenon of the Deardorff flux qualitatively correctly, it
is quite possible that the nature of the underlying coherent structures
may require additional refinements.

\subsection{Depth dependence of the Deardorff term}
\label{DepthDependence}

To estimate the resulting depth dependence of the Deardorff term in
\Eq{nabD}, we must know how the $\overline{s^2}$ associated with
the Deardorff term varies with depth.
If the entropy rain was purely of the form of Hill vortex-like structures,
as discussed above, their filling factor $f_s$ would decrease with
increasing depth and density like
\EQ
f_s\propto\meanrho^{\,-\zeta},
\label{fsDef}
\EN
where $\zeta=0.8$ has been found for spherical vortex structures
descending in an isothermally stratified layer; see \App{HillVortices}.
However, \Eq{fsDef} is not contingent on Hill vortices,
but is a consequence of downdrafts along a density gradient.
For purely spherical compression one would expect $\zeta=2/3$,
while for horizontal compression one has $\zeta=1$.

If we neglect non-ideal (radiative or viscous) effects, as well as
entrainment between up- and downflows, the difference
$\Delta S=S_\uparrow-S_\downarrow$ between the entropy of the slowly
rising surroundings, $S_\uparrow$, and that in the downward propagating
vortex, $S_\downarrow$, would remain constant and equal to the entropy
deficit $\Delta S_0$ suffered by overturning motions at the surface,
which is the only location where radiative losses are significant.
On sufficiently short length scales, however, radiative heating from the
surroundings would erode this entropy difference and lead to a decrease
of the effective $\Delta S$.
We model this by considering $\zeta$ an adjustable parameter that is
{\em increased} relative to the value $0.8$ that we expect in
the absence of radiation, i.e., $f_s$ decreases more strongly.
The resulting fractional entropy difference, $f_s\,\Delta S_0$,
therefore decreases faster with depth than for $\zeta=0.8$.

As will become clear from the considerations below, 
if there is no entrainment from the upflows into the downdrafts
and all of the downflows were confined to the small surface area of the
vortex structures, the resulting downward-directed {\em kinetic} energy
flux would increase with depth and eventually exceed the enthalpy flux.
This would be unphysical, because convection should lead to outward
energy transport.
However, if there is entrainment with associated mixing, the actual
filling factor (fractional area) of downflows, which we denote
by $f$, would be larger and the {\em mean} entropy difference
$\Delta\meanS$ would be diluted correspondingly such that
\EQ
f\,\Delta\meanS=f_s\,\Delta S_0\quad
\left(\propto\meanrho^{\,-\zeta}\,\Delta S_0\right).
\label{fsDef}
\EN
We recall that non-ideal effects can be captured by choosing
$\zeta>0.8$ in the prescription \eq{fsDef}.
The physics of entrainment has been modeled in the stellar context by
\cite{RZ95} and the extent of entrainment has been quantified in
the realistic surface simulations by \cite{TS11}, who determined the
entrainment length scale as the typical scale height of the mass flux
in the up- and downflows separately.
They call this scale the mass mixing length and find its value to be
comparable to $\HP$.

To estimate the depth dependence of various horizontal averages, and
to compute enthalpy and kinetic energy fluxes, we now compute various
averages in the two-stream approximation, in which all horizontal averages
are the sum of the fraction $1-f$ of the value in upflows and the fraction
$f$ of the value in downflows.
Thus, for the mean specific entropy stratification we have
\EQ
\meanS=(1-f)\meanS_\uparrow+f\meanS_\downarrow=\meanS_\uparrow-f\,\Delta\meanS,
\label{meanSfill}
\EN
where $\meanS_\uparrow$ and $\meanS_\downarrow$
are the mean specific entropies in up- and downflows, and
$\Delta\meanS=\meanS_\uparrow-\meanS_\downarrow$ is their difference.
We expect that $\meanS_\uparrow\approx S_\uparrow$ will be
constant if there is no heating in the upwellings, while
$\meanS_\downarrow\approx(1-f_s/f)S_\uparrow+(f_s/f)S_\downarrow$ will
be dominated by contributions from $S_\uparrow$ due to entrainment.

To calculate $\overline{s^2}$, we must first compute the fluctuating
quantity $s=S-\meanS$ and then average the squared values in up- and
downflows, which, using \Eq{meanSfill}, gives
\EQ
\overline{s^2}=(1-f)\,(\meanS_\uparrow-\meanS)^2
+f\,(\meanS_\downarrow-\meanS)^2=\hat{f}\,(\Delta\meanS)^2,
\label{s2m}
\EN
where $\hat{f}=(1-f)f$ has been introduced as a shorthand.
Analogously to the specific entropy, we find for the velocity
\EQ
\meanU_z=(1-f)\meanU_\uparrow+f\meanU_\downarrow
=\meanU_\uparrow-f\,\Delta\meanU,
\EN
where $\meanU_\uparrow>0$ and $\meanU_\downarrow<0$ are the mean up- and
downflow velocities with $\Delta\meanU=\meanU_\uparrow-\meanU_\downarrow$.
Since the densities in up- and downflows are nearly the same, especially in
deeper layers, we have $\meanU_z\approx0$ from mass conservation.
With this, we find
\EQ
\urms^2\equiv\overline{u_z^2}=(1-f)\,\meanU_\uparrow^2
+f\,\meanU_\downarrow^2=\hat{f}\,(\Delta\meanU)^2,
\label{u2m}
\EN
and thus $\meanU_\uparrow/\urms=[f/(1-f)]^{1/2}$.
Similarly, we calculate
\EQ
\overline{u_z^3}=(1-f)\,\meanU_\uparrow^3+f\,\meanU_\downarrow^3
=-\hat{f}\,(1-2f)\,(\Delta\meanU)^3,
\EN
which is negative for $f<1/2$, and finally
\EQ
\overline{u_z s}=(1-f)\,\meanU_\uparrow\meanS_\uparrow
+f\,\meanU_\downarrow\meanS_\downarrow=\hat{f}\,\Delta\meanU\,\Delta\meanS.
\label{uzs_eqn}
\EN
Thus, the kinetic energy flux, $\overline{\rho\uu^2 u_z}/2$, which
reduces to $\meanrho\,\overline{u_z^3}/2$ in this two-stream approximation,
is given by
\EQ
F_{\rm kin}=-\phi_{\rm kin}\,\meanrho\urms^3,
\label{Fkin}
\EN
where $\phi_{\rm kin}=(1/2-f)/\hat{f}^{1/2}$ is a positive prefactor
(corresponding to downward kinetic energy flux) if $f<1/2$.
\cite{SNGBS09} find $f\approx1/3$, nearly independently of depth,
which yields $\phi_{\rm kin}\approx\sqrt{2}/4\approx0.35$; see
\Tab{SummaryPHIKIN}, where we list $\phi_{\rm kin}$ and
$-\meanU_\downarrow/\urms=[(1-f)/f]^{{1/2}}$ for selected values of $f$.
The enthalpy flux is proportional to $\overline{u_z s}$ and, using
\Eqs{Fenth}{uzs_eqn} together with \Eqs{s2m}{u2m}, we find
$F_{\rm enth}=\meanrho\, \meanT\,\urms\srms$.
In \App{DeltaS_Ma2} we show that
$\meanT\srms=\urms^2\kf \HP/(a_{\rm MLT}\nabad)$, where $a_{\rm MLT}=1/8$
is a geometric factor in standard MLT.\footnote{Not to be confused with
the parameter $\alpha_{\rm mix}$ of \Sec{RelaxationTime} below.}
This leads to
\EQ
F_{\rm enth}=\phi_{\rm enth}\,\meanrho\urms^3
\label{Fenth2}
\EN
with $\phi_{\rm enth}=\kf\HP/(a_{\rm MLT}\nabad)$.
This yields $\phi_{\rm enth}\approx20$, which is rather large.
By contrast, \cite{BCNS05} determined a quantity $k_u$ such that
$\phi_{\rm enth}=k_u^{-3/2}\approx4$.

We see that the presence of a kinetic energy flux just
modifies the usual expression for the convective flux,
which then becomes the sum of enthalpy and kinetic energy fluxes;
see \Sec{Derivation}.
This is compatible with recent simulations of R.\ F.\ Stein
(2016, private communication), in which the fractional kinetic energy flux
increaes toward the deeper parts ($\ga40\Mm$) of the domain.

\begin{table}[t!]\caption{
$\phi_{\rm kin}$, $-\meanU_\downarrow/\urms$, and $\meanU_\uparrow/\urms$
for selected values of $f$.
}\centerline{\begin{tabular}{ccccccc}
$f$                        & 1/2 & 1/3  & 0.14  & 0.015 & 0.0006\\
\hline
$\phi_{\rm kin}$           &  0  & 0.35 &   1   &   4   &   20 \\
$-\meanU_\downarrow/\urms$ &  1  & 1.4  &  2.5  &   8   &   40 \\
$\meanU_\uparrow/\urms$    &  1  & 0.7  &  0.4  &  0.12 &  0.025
\label{SummaryPHIKIN}\end{tabular}}\end{table}

When $f$ becomes small ($<0.14$), $\phi_{\rm kin}$ exceeds unity and for
$f<0.015$, $\phi_{\rm kin}$ exceeds the estimate $\phi_{\rm enth}\approx4$
found by \cite{BCNS05}, so the sum of enthalpy and kinetic energy fluxes
may become negative, which appears unphysical.
In that case, the idea of reconciling the results of
\cite{Hanasoge} with such convection transporting the solar luminosity
could be problematic, unless both up- and downflows were to occur on
sufficiently small scales.
In the case $f=0.14$, the resulting
$\meanU_\downarrow$ would still not be particularly fast and only 2.5 times
larger than $\urms$; for $f=0.015$ we have $-\meanU_\downarrow\approx8\urms$;
see \Tab{SummaryPHIKIN}.
Conversely, if the cold entropy blobs were much smaller and were still
to contribute significantly to the total energy flux, i.e., if they were
much faster, this might not be compatible with an upward directed total
energy transport.
A possible alternative might be suggested by the compressible simulations
of \cite{Cat91}, where the downward-directed kinetic energy flux in the
downdrafts was found to balance the upward directed enthalpy flux in
the downdrafts.
This would imply that convection would transport energy only in the
upwellings.
This result has however not been confirmed in realistic surface
simulations \citep{SBN92}.

\EEq{fsDef} shows how $f_s$ decreases with depth, but its actual
value and that of $\nabD$ remains undetermined.
In the following, we propose a quantitative prescription for $\nabD$
by estimating its value within the top few hundred kilometers.
In those top layers, we expect $\nabD$ to reach its maximum value,
$\nabD^{\max}$, and it should be a certain fraction of $\nabla-\nabad$.
In \App{SurfaceValue} we show that
$\nabD^{\max}/(\nabla-\nabad)_{\max}\approx3$.

In deeper layers, where the local value of $\nabla-\nabad$ has become
small, $\nabD$ should scale with $\overline{s^2}=\hat{f}\,(\Delta\meanS)^2$,
and therefore, using \Eq{fsDef}, it should be proportional to
$f_s^2(z)\propto\meanrho^{\,-2\zeta}$, in addition to an $\Ma^{-2}$ factor;
see \Eq{nabD}.
Thus, we have
\EQ
\nabD=f_s^2\,\nabD^{\max},
\label{nabD_ansatz}
\EN
where we have used for $f_s$ the expression
\EQ
f_s=f_{s0} (\meanrho/\rho_\ast)^{-\tilde\zeta}
\quad\mbox{(for $\meanrho>\rho_\ast)$}.
\label{fzeta}
\EN
Here, $f_{s0}$ is a prefactor determining the strength of the
resulting Deardorff flux and $\rho_\ast$ is the density at the point
at which $\nabla-\nabad$ is equal to its maximum value
(which is just below the photosphere).
The new exponent $\tilde\zeta=\zeta-\Delta\zeta$ takes the scaling of
Mach number with density into account, i.e.,
\EQ
\Ma\propto\meanrho^{\,-\Delta\zeta},
\label{Madep}
\EN
where $\Delta\zeta$ ($>0$) will be computed in \Sec{SuperadiabaticGradient}.

\subsection{Kinetic energy flux}

The importance of the kinetic energy flux has been stressed for some
time as a property of compressible stratified convection \citep{HTM84,Cat91}.
This flux is related to the asymmetry of up- and downflows \citep{SN89}
and is non-vanishing when $f\neq1/2$; see \Sec{DepthDependence}.
It is neglected in standard MLT, as has been discussed by \cite{Arnett15}.
Just like the Deardorff flux, the kinetic energy flux is also a
non-gradient flux, but it is always directed downward for $f<1/2$
and must therefore be overcome by the enthalpy flux so that energy
can still be transported outwards.
However, since the lowest order correlations in $F_{\rm kin}$
are triple correlations, there is no $\tau$ approximation treatment
analogous to that of the Deardorff flux.
However, by comparing $F_{\rm kin}=-\phi_{\rm kin}\,\meanrho\urms^3$
with the enthalpy flux in \Eq{Fconv_urms2}, it is possible to define
a corresponding nabla term via $F_{\rm kin}=-\onethird\meanrho
\cP \meanT \, (\taured\urms^2/\HP)\nabkin$.
This yields
\EQ
\nabkin=3\,\phi_{\rm kin}\,\urms \HP/\taured\cP\meanT,
\EN
which is obtained analogously to $\nabD$ in \Eq{nabD}.
The prefactor is here defined with a positive sign, so
the total non-radiative flux caused by the turbulence is
proportional to $\nabla-\nabad+\nabD-\nabkin$.

Obviously, if $\nabkin$ is strictly proportional to $\nabla-\nabad$,
the addition of the $\nabkin$ term does not modify standard MLT,
provided that $\nabkin$ depends just on the {\em local} value of
$\urms$, which in turn is, again, assumed to depend just on the local
value of $\nabla-\nabad$.
This is different for $\nabD$, which depends on the
entropy deficit produced by cooling near the surface and in this way on
the value of $\nabla-\nabad$ at the position where $\meanrho=\rho_\ast$.
The $\nabD$ term is therefore truly nonlocal.
In the following, we combine kinetic and enthalpy fluxes
into a total contribution $F_{\rm conv}=F_{\rm enth}+F_{\rm kin}$ which
arises from convection.

\section{Modified mixing length model}
\label{Modified}

\subsection{Stratification and flux balance}

To construct an equilibrium model, we begin by considering first the
case without convection, so the flux $F$ is carried by radiation alone.
Hydrostatic and thermal equilibrium then imply $\dd\meanP/\dd z=-\meanrho g$
and $K\dd\meanT/\dd z=-F$, or, alternatively for the logarithmic gradients,
\EQA
\dd\ln \meanP/\dd z&=&-\meanrho g/\meanP,
\label{dlnP}
\\
\dd\ln \meanT/\dd z&=&-F/(K\meanT).
\label{dlnT}
\ENA
The double-logarithmic temperature gradient is obtained by dividing
the two equations through each other, i.e.,
\EQ
\nabla={\dd\ln \meanT\over\dd\ln \meanP}
={F\meanP\over K\meanT\,\meanrho g}={F\cP\nabad \over Kg},
\label{nabla}
\EN
where we have used the perfect gas equation of state in the form
$\meanP/\meanT\,\meanrho=\cP-\cV=\cP\,(1-1/\gamma)=\cP\nabad$.
If the energy is no longer carried by radiation alone, $\nabla$ cannot be
computed from \Eq{nabla}, but we have to invoke a suitable theory of
convection.
In standard MLT, one obtains $\nabla$ as a solution of a cubic equation
\citep{Vit53,KW90}.
In the following, we consider a modification that accounts for the
possibility of a Deardorff flux.

Flux balance implies that the sum of the radiative, enthalpy, and kinetic
energy fluxes equals the total flux, i.e.,
\EQ
F_{\rm tot}=F_{\rm rad}+F_{\rm conv}.
\label{FluxBalance}
\EN
where $F_{\rm conv}=F_{\rm enth}+F_{\rm kin}$ is the flux that arises
from convection.
In MLT, it is customary to express these fluxes
in terms of nablas.
For a given double-logarithmic temperature gradient $\nabla$,
the radiative flux is evidently
\EQ
F_{\rm rad}={Kg\over \cP\nabad}\,\nabla,
\EN
where $\nabla$ characterizes the {\em actual} temperature gradient.
We can also define a {\em hypothetical} radiative temperature gradient
$\nabrad$ that would result if all the energy were carried by radiation,
so we can write 
\EQ
F_{\rm tot}={Kg\over \cP\nabad}\,\nabrad,
\label{nabrad_Def}
\EN
which follows from \Eq{nabla}.
We note in passing that the ratio between $F_{\rm tot}$ and the radiative
flux carried by the adiabatic temperature gradient, $Kg/\cP$, is known
in laboratory and theoretical studies of convection as the Nusselt number
\citep{HTM84}, whose local value is thus equal to
\EQ
\mbox{Nu}=\nabrad/\nabad.
\label{Nu}
\EN
Finally, as explained in \Sec{Deardorff}, we have from \Eq{Fconv_urms2}
\EQ
F_{\rm enth}=F_0\left(\nabla-\nabad+\nabD\right),
\label{Fconv_urms1}
\EN
with $F_0=\onethird\meanrho \cP \meanT \taured\urms^2/\HP$,
and $F_{\rm kin}$ being essentially proportional to $F_{\rm enth}$;
cf.\ \Eqs{Fkin}{Fenth2}.
Flux equilibrium then implies
\EQ
\nabrad=\nabla+\epsilon\left(\nabla-\nabad+\nabD\right),
\label{epsilon_eqn}
\EN
where $\epsilon=F_0 \cP\nabad(1-\phi_{\rm kin}/\phi_{\rm enth})/Kg$.
However, the expression for $F_0$ involves the still unknown
values of $\urms$ and $\taured$, which will be discussed in
\Sec{RelaxationTime}.

\begin{figure*}[t!]\begin{center}
\includegraphics[width=\textwidth]{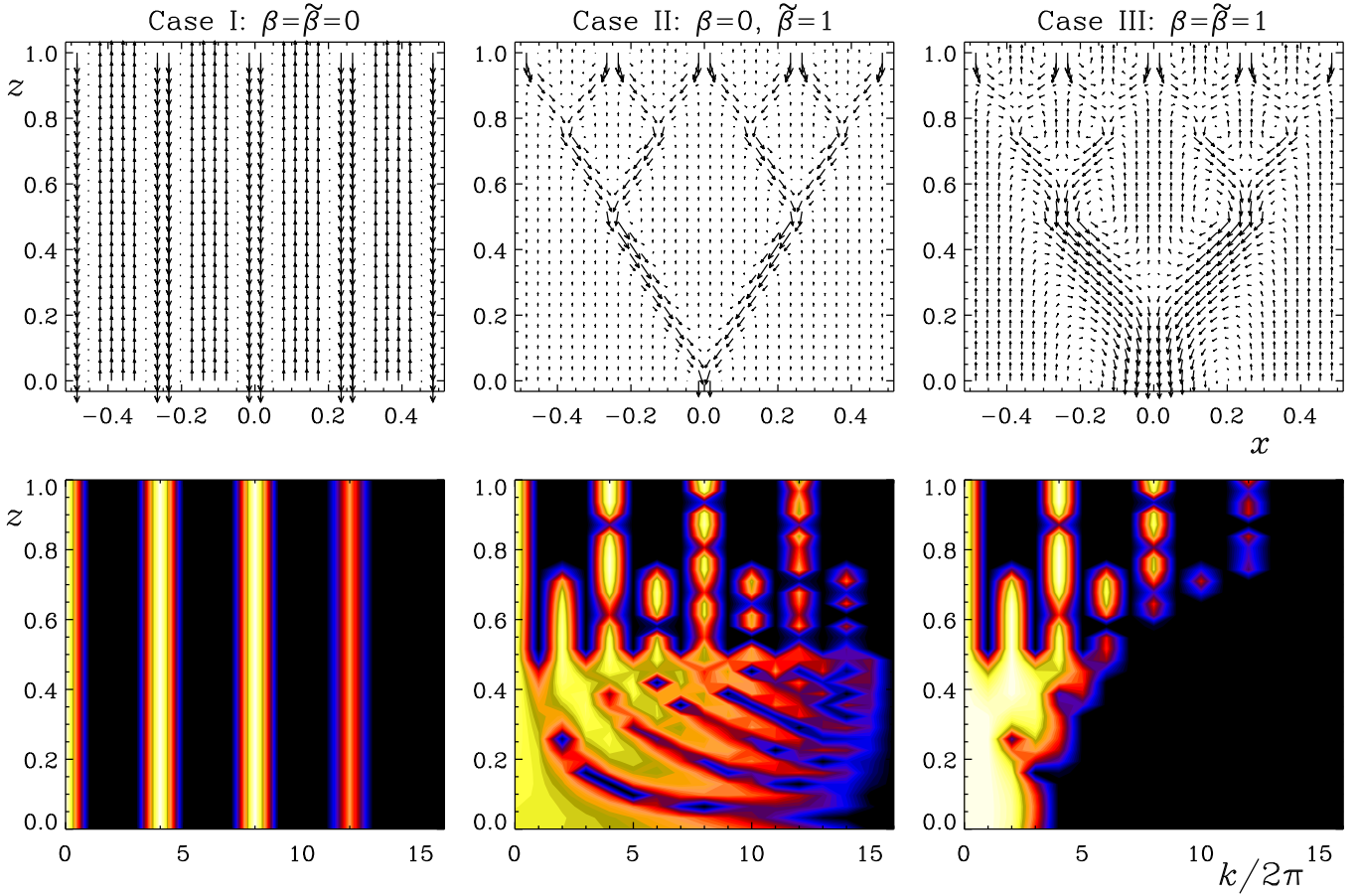}
\end{center}\caption[]{
Illustrative flow structures (upper row) and corresponding
horizontal power spectra (lower row) associated with the three
combinations of $\beta$ and $\tbeta$ considered in this paper.
Light shades correspond to large logarithmic power, which is seen to
extend over large values of $k$ for $\beta=0$ (Cases~I and II)
and is confined to progressively smaller $k$ at larger depths
when $\beta=1$ (Case~III).
}\label{pbetacases2}\end{figure*}

\subsection{Relaxation time and mixing length}
\label{RelaxationTime}

In convection the relevant time scale is the turnover time,
which we write as $\tau=1/\urms\kf$.
We argue that $\kf$ should be estimated via the separation
between the entropy rain structures and not their thickness.
This becomes important in the picture in which the downdraft threads
of the entropy rain merge with neighboring ones to form a tree-like
structure, as seen in the surface simulations \citep{SN89,Spr97},
leading therefore to different scalings of the two length scales
(separation and thickness of structures) with depth.
It is then not obvious which of these scales are more relevant
for determining the $\kf$ that is relevant for mixing.
In view of the uncertainty regarding the choice of $\kf$, as well as
for comparison with standard MLT, we consider models with a fixed
value of $\kf$ as well as the more conventional case in which
$\kf \HP\approx\const$.
To capture the various cases in one expression, we assume in the following
\EQ
\kf=\alpha_{\rm mix}^{\beta}\HP^{-1} (\kfz\HP)^{1-\beta},
\label{kfbeta}
\EN
where $\beta=0$ corresponds to $\kf=\kfz$ with a fixed value $\kfz$
of the wavenumber of the entropy rain and $\beta=1$ corresponds to
$\kf\HP=\alpha_{\rm mix}$, where we have allowed for the possibility
of a free mixing length parameter, $\alpha_{\rm mix}$, as is commonly
done in standard MLT.
It is not to be confused with the parameter $a_{\rm MLT}$ that
was introduced in \Sec{DepthDependence} and \App{DeltaS_Ma2}.
Negative values of $\beta$ would correspond to shrinking scales,
but will not be considered here.
Likewise, non-integer values of $\beta$ are conceivable,
but will also not be considered here.

Returning now to the discussion of the relaxation time $\tau$ in the beginning
of this subsection, instead of associating it with the turnover time
$(\urms\kf)^{-1}$, we will allow for the possibility of an additional
dilution factor $\phi(z)$ and write $\tau=(\urms\kf\phi)^{-1}$.
Here, $\phi(z)$ increases with depth in a similar fashion as the
scale height, so we assume, in analogy with our treatment of $\kf$ in
\Eq{kfbeta}, the expression $\phi=(\kfz\HP/\alpha_{\rm mix})^{\beta-\tbeta}$.
This dilution factor only enters in expressions involving the turbulent
diffusivity, such as in \Eq{FG}, and hence terms involving $\tau$ or $\taured$.
For $\beta=1$, the case $\tbeta=1$ corresponds to the usual MLT concept,
while $\tbeta=0$ corresponds to a value of $\phi$ that increases with depth.

In \Fig{pbetacases2} we present illustrative flow structures
as well as their depth-dependent horizontal power spectra associated
with the three combinations of $\beta$ and $\tbeta$ considered here.
At the top of the domain, the size and separation of flow structures are the
same in all three cases.
They remain constant with depth in the case $\beta=\tbeta=0$ (Case~I),
while for $\beta=0$ and $\tbeta=1$ (Case~II), the separation increases
with depth, but the thickness is still constant.
Finally, for $\beta=\tbeta=1$ (Case~III), both thickness and separation
increase with depth.
The filling factor decreases with depth in the case $\beta=0$ and
$\tbeta=1$, which could potentially make the kinetic energy flux divergent
with depth, while for both $\beta=\tbeta=0$ and $\beta=\tbeta=1$,
the filling factor is independent of height.

With these preparations in place, we can write $F_0$ in \Eq{Fconv_urms1}
in the form
\EQ
F_0=(\sigma/3\gamma\nabad) \, \meanrho \urms \cs^2
\alpha_{\rm mix}^{-\tbeta} (\kfz \HP)^{-(1-\tbeta)},
\label{F0}
\EN
where
\EQ
\sigma\equiv\taured\urms\kf\phi=\urms/(\urms+\iota c_\gamma/\phi)
\label{urms}
\EN
quantifies the radiative heat exchange between convective elements
and the surroundings.
The $\iota$ term defined in \Eq{taucool}, has a maximum at
$\ell\kf=\sqrt{3}$, which typically occurs near the surface
\citep[e.g.][]{BB14}.

In \Eq{F0}, the local value of the turbulent rms velocity always depends
on the actual flux transported, and therefore it must also depend on
$\nabla-\nabad+\nabD$.
On dimensional grounds, since $F_{\rm enth}$ is proportional to $\urms^3$,
we have
\EQ
\urms=c_0\left(\nabla-\nabad+\nabD\right)^{1/2},
\label{prefactor}
\EN
so $F_{\rm enth}$ and $F_{\rm kin}$ are proportional to
$(\nabla-\nabad+\nabD)^{3/2}$.
Standard mixing length arguments can be used to show that
the prefactor in \Eq{prefactor} is a fraction of $\cs$.
As shown in \App{App}, we have
\EQ
c_0/\cs=\sqrt{\sigma a_{\rm MLT}/3\gamma}\,
\alpha_{\rm mix}^{-\beta'} (\kfz \HP)^{-(1-\beta')},
\label{urms0}
\EN
where $\beta'=(\beta+\tbeta)/2$.
It is then clear that $F_{\rm kin}$ scales with $\HP$ like
$(\kfz \HP)^{-3(1-\beta')}$.
This is not the case for $F_{\rm enth}$, however, because of
the factor $(\kfz \HP)^{1-\tbeta}$ which, together with the
$(\kfz \HP)^{1-\beta'}$ factor, gives the scaling proportional
to $(\kfz \HP)^{2(1-\beta'')}$, where $\beta''=(\beta'+\tbeta)/2$.

\subsection{Equation for the superadiabatic gradient}
\label{SuperadiabaticGradient}

Now that we know $F_0$, we can solve the equation for the superadiabatic
gradient.
This leads to an equation that is similar to the cubic equation for $\nabla$,
which is familiar from standard MLT \citep{KW90},
\EQ
\nabrad=\nabla+\epsilon_\ast\left(\nabla-\nabad+\nabD\right)^\xi,
\label{nabla_final}
\EN
where $\xi=3/2$, and $\epsilon_\ast=(1-\phi_{\rm kin}/\phi_{\rm enth})
(\epsilon_{\rm enth}+\epsilon_{\rm kin})$
has contributions from the enthalpy and kinetic energy fluxes.
These expressions are similar to $\epsilon$ in \Eq{epsilon_eqn}, except that
$\epsilon_{\rm enth}$ is evaluated with $c_0$ in place of $\urms$, i.e.,
\EQ
\epsilon_{\rm enth}=(\sigma/3\gamma)^{3/2}\,a_{\rm MLT}^{1/2}\;(\cs^3/\chi g)
\,\alpha_{\rm mix}^{-2\beta''} (\kfz\HP)^{-2(1-\beta'')},
\label{eps2}
\EN
where $\chi=K/\meanrho \cP$ is the radiative diffusivity.
This shows that $\epsilon_\ast$ is essentially a P\'eclet number
based on $\cs$.
The contribution from the kinetic energy flux is given by
\EQ
\epsilon_{\rm kin}/\epsilon_{\rm enth}=-\phi_{\rm kin}
a_{\rm MLT}\nabad \, \alpha_{\rm mix}^{-\beta} (\kfz\HP)^{-(1-\beta)}.
\label{eps3}
\EN

We note in passing that $\epsilon_\ast$ is also related to the
Rayleigh number, which is commonly defined in laboratory and numerical
studies of convection; see \App{Rayleigh}.
Furthermore, because of convection and the resulting bulk mixing,
$\meanS$ is now approximately constant, and therefore, unlike in the
non-convecting reference solution with $K=\const$
(\Sec{HighlyUnstable}), $K$ can no longer be constant,
but it reaches a minimum at the point where
$\kappa$ is maximum, which turns out to be at a depth of about $1\Mm$
in the convection models presented below.
Since $\epsilon_\ast$ is inversely proportional to $K$, it reaches
a maximum at that depth and falls off both toward the top and the
bottom of the convection zone.

An essential difference between \Eq{nabla_final} and the usual one in
MLT is the presence of $\nabD$ arising from the Deardorff flux.
Within the usual MLT, where $\nabD=0$, one finds that $\nabla$ is slightly
above $\nabad$, but now it might instead be slightly above
$\nabad-\nabD$.
There are indeed two possibilities for convecting solutions
($F_{\rm conv}>0$), one corresponding to a Schwarzschild-stable solution,
\EQ
\nabad-\nabD < \nabla < \nabad \quad\mbox{(stable)},
\label{stable}
\EN
and one that is Schwarzschild unstable,
\EQ
\nabad-\nabD < \nabad < \nabla \quad\mbox{(unstable)}.
\label{unstable}
\EN
Which of the two possibilities is attained depends on the value of
$\nabD$ and also on details of the solution.
As will be discussed in \Sec{UnstablyStratified} below, entropy rain
convection may actually still be Schwarzschild unstable without
exciting giant cell convection if
small-scale turbulent viscosity and diffusivity are strong enough
so that the local turbulent Rayleigh number for the deeper layers
is subcritical.

In this connection, we note that in standard MLT, one includes the
effects of radiative cooling of the convective elements
in a different manner than here.
Instead of $\nabla-\nabad$, the effective buoyancy force is written as
$\nabla-\nabla'$, where $\nabla'$ always lies between $\nabla$
and $\nabad$ \citep{Vit53}.
Thus, one has $\nabad < \nabla' < \nabla$, which resembles
\Eq{unstable} with a negative value of $\nabD$.

\begin{figure}[t!]\begin{center}
\includegraphics[width=\columnwidth]{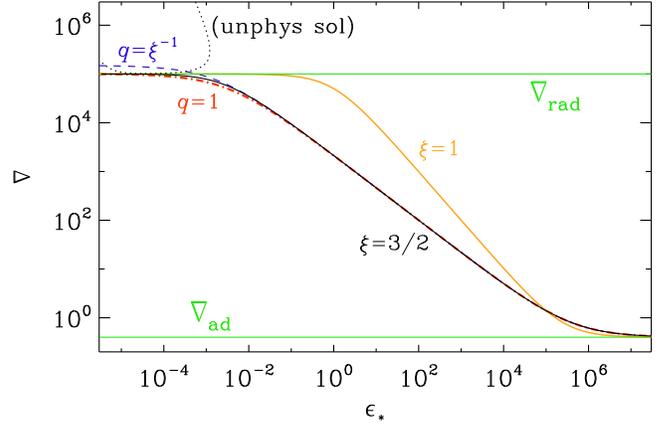}
\end{center}\caption[]{
Solution of \Eq{nabla_final} for $\xi=3/2$ and $\nabrad=10^5$
(solid black), compared with the approximation \eq{DeltaNabla}
for $q=\xi^{-1}$ (dashed blue) and $q=1$ (dash--dotted red),
and $\xi=1$ (thin orange).
The dotted line gives an additional unphysical solution
of \Eq{nabla_final} for $\xi=3/2$.
The limiting cases $\nabla=\nabad$ and $\nabrad$ are shown as
thin horizontal green lines.
}\label{pcubic_single}\end{figure}

To understand the nature of the solutions of \Eq{nabla_final},
it is instructive to treat $\xi$ as an adjustable parameter.
For given values of $\nabrad$ and $\tnabad\equiv\nabad-\nabD$,
the case $\xi=1$ yields
\EQ
\nabla(\epsilon_\ast)={\nabrad+\epsilon_\ast\tnabad\over1+\epsilon_\ast}.
\label{xi1}
\EN
It shows that $\nabla\to\nabrad$ for $\epsilon_\ast\to0$ (stable surface
layers) and $\nabla\to\tnabad$ for $\epsilon_\ast\gg1$ (deeper layers).
Next, to discuss the general case $\xi\neq1$, we define
$\Delta\nabla=\nabla-\tnabad$ and $\Delta\nabrad=\nabrad-\tnabad$.
For $\Delta\nabrad<0$ we have $\Delta\nabla=\Delta\nabrad$, while for
$\Delta\nabrad>0$ and $\Delta\nabla\ll1$, a useful approximation is
\EQ
\Delta\nabla\approx\Delta\nabrad^{1/\xi}\left/\left(q
\Delta\nabrad^{1/\xi-1}+\epsilon_\ast^{1/\xi}\right)\right.,
\label{DeltaNabla}
\EN
where $q=\xi^{-1}$.
It agrees with \Eq{xi1} in the special case $\xi=1$, where
$\Delta\nabla=\Delta\nabrad/(1+\epsilon_\ast)$.
In \Fig{pcubic_single} we plot $\nabla$ versus $\epsilon_\ast$
for $\nabrad=10^5$.
The approximation yields $\nabla>\nabrad$ for $\epsilon_\ast<10^{-3}$,
which is unphysical.
This can be mitigated by choosing $q=1$; see \Fig{pcubic_single}.

For the relevant case of large values of $\epsilon_\ast$, we have
\EQ
\Delta\nabla\approx(\nabrad/\epsilon_\ast)^{1/\xi}.
\label{DelNabApprox}
\EN
This relation is useful because, even though both
$\nabrad$ and $\epsilon_\ast$ depend on $K$, their ratio does not
and is given by
\EQ
{\nabrad\over\epsilon_\ast}
={3 \nabad \over \sigma} \, {F_{\rm tot}/\meanrho \over c_0 g/\kf}
\propto {(\kfz \HP)^{2-2\beta''} \over \meanrho \cs^3 }
\propto\meanT^{\,-(1+2\beta'')},
\label{epsratio}
\EN
where $\beta''=(\beta+3\tbeta)/4$.
Thus, $\Delta\nabla\propto\meanT^{\,-(1+2\beta'')/\xi}$ gives the
scaling of $\Delta\nabla$ for the bulk of the convection zone.
Therefore, looking at \Eq{prefactor}, we find
$\Ma\propto\meanT^{\,-m}$ with
\EQ
m=1-\beta'+(1/2+\beta'')/\xi.
\label{mexpression}
\EN
For $\xi=3/2$, using the relation $3\beta'-2\beta''=\beta$,
we find that $m=(4-\beta)/3$ is independent of the value of $\tbeta$.
Thus, we have $m=4/3$ with $\beta=0$ and $m=1$ with $\beta=1$.
For isentropic stratification, this implies for the Mach number,
given by \Eq{Madep}, the following scaling: $\Delta\zeta=8/9$ with $\beta=0$
and $\Delta\zeta=2/3$ with $\beta=1$.

\subsection{Relative importance of the Deardorff term}
\label{RelativeImportance}

In \Sec{DepthDependence} we have considered the depth dependence
of the Deardorff term via \Eqs{nabD_ansatz}{fzeta}.
However, for $\nabD$ to be important at increasing depths, it must
exceed the {\em sub}adiabatic gradient $\nabad-\nabla$, because otherwise
it would not be possible for the Deardorff term to make
$\nabla-\nabad+\nabD$ positive.
From \Eqs{DelNabApprox}{epsratio} we see that $\Delta\nabla$ depends
on $\meanT$ in a power-law fashion.
We are particularly interested in the conditions under which
$\Delta\nabla$ falls off faster than $\nabD$, because that would
ensure that the $\nabD$ term remains important even at larger depths.

\begin{table}[t!]\caption{
Comparison of $\zeta_{\Delta\nabla}$, $2\tilde\zeta=2\zeta-2\Delta\zeta$,
and their difference for various combinations of $\beta$ and $\tbeta$ 
using $\zeta=1$.
}\centerline{\begin{tabular}{lccccccc}
& $\beta$ & $\tbeta$ & $\beta'$ & $\beta''$ &
$\zeta_{\Delta\nabla}$
& $2\tilde\zeta$ & $\!\!\zeta_{\Delta\nabla}-2\tilde\zeta$ \\
\hline
Case~I   & 0 & 0 &  0  &  0  &  4/9 & 2/9 & 2/9 \\
Case~II  & 0 & 1 & 1/2 & 3/4 & 10/9 & 2/9 & 8/9 \\
         & 1 & 0 & 1/2 & 1/4 &  2/3 & 2/3 &  0  \\
Case~III & 1 & 1 &  1  &  1  &  4/3 & 2/3 & 2/3 \\
\label{SummaryBETA}\end{tabular}}\end{table}

For $\xi=3/2$, and since the convection zone is nearly isentropically
stratified ($\meanT\propto\meanrho^{\,2/3}$), we have
\EQ
\Delta\nabla\propto\meanrho^{\,-\zeta_{\Delta\nabla}}
\quad\mbox{with}\quad \zeta_{\Delta\nabla}={4\over9}(1+2\beta'').
\EN
In \Tab{SummaryBETA} we compare for various combinations of $\beta$
and $\tbeta$ the exponents for $\Delta\nabla$ and $\nabD$, i.e.,
$2\tilde\zeta=2\zeta$ and $2\Delta\zeta$, where $\Delta\zeta=2m/3$.
In all cases we have chosen $\zeta=1$, i.e., we allow for moderately
non-ideal (radiative) effects relative to the ideal case with $\zeta=0.8$.
We see that the difference is positive and non-vanishing in all
cases, except for $\beta=1$ and $\tbeta=0$.
In the following, we present solutions for all of the remaining three cases.
Before doing this, let us recapitulate what led to the threefold
dominance of $\tbeta$ over $\beta$.
For better illustration, we summarize in \Tab{Summary} the various
relationships that led to the scaling of $\Delta\nabla$ with $\kfz\HP$.

In the first expression for $F_{\rm enth}$, $\beta$ enters because it
characterizes the relation between the buoyancy force proportional to
$\delta T/\meanT$ and advection proportional to $\kf\urms^2$; see
\App{DeltaS_Ma2}.
For thin threads, we expect the relevant $\kf$ to be large, i.e., $\beta=0$
(Cases~I and II in \Fig{pbetacases2}).
Next, in the second expression for $F_{\rm enth}$, we have used a mean-field
expression to relate $F_{\rm enth}$ to $\Delta\nabla$ via a turbulent
diffusivity proportional to $\tau\urms^2\approx\urms/\kf\phi$, where,
as argued above, the dilution factor $\phi$ has entered.
It is this expression that is closest to conventional MLT, because here
we expect $\tbeta=1$.

\begin{table}[t!]\caption{
Illustration of scaling relationships with $\kfz\HP$.
}\centerline{\begin{tabular}{ll}
\hline
\hline
$F_{\rm enth}\propto\urms^3\,(\kfz\HP)^{1-\beta}$ &
  $\!\!$buoyancy force\\
$F_{\rm enth}\propto\urms\Delta\nabla\,/\,(\kfz\HP)^{1-\tbeta}$   &
  $\!\!$mean-field expression \\
$\urms\;   \propto(\Delta\nabla)^{1/2}/(\kfz\HP)^{1-\beta'}$ &
  $\!\!$$\beta'\;=(\beta\!+\!\tbeta)/2$ \\
$F_{\rm enth}\propto(\Delta\nabla)^{3/2}/(\kfz\HP)^{2(1-\beta'')}$ &
  $\!\!$$\beta''=(\beta'\!+\!\tbeta)/2=(\beta\!+\!3\tbeta)/4 $ \\
$F_{\rm kin}\;\;\propto(\Delta\nabla)^{3/2}/(\kfz\HP)^{3(1-\beta')}$ &
  $\!\!$kinetic energy flux \\
\hline
\label{Summary}\end{tabular}}\end{table}

The remaining two relationships in \Tab{Summary} explain why the $\tbeta$
term appears three times more dominantly than the $\beta$ term.
By equating the first two expressions for $F_{\rm enth}$ in \Tab{Summary},
we find first of all the relation between $\urms$ and $\Delta\nabla$,
where $\beta$ and $\tbeta$ contribute with equal shares through
$\beta'=(\beta+\tbeta)/2$.
However, to see the scaling of $\Delta\nabla$, we need to go back to
the second expression for $F_{\rm enth}$, because it changes only weakly
with depth.
Now, $\beta'$ and $\tbeta$ contribute with equal shares, and this means
that $\tbeta$ has now become three times more dominant than $\beta$ through
the expression $\beta''=(\beta+3\tbeta)/4$.
Thus, for $\beta=0$ and $\tbeta=1$, $\Delta\nabla$ shows nearly
the standard scaling with $\HP$.
Furthermore, looking again at the first expression for $F_{\rm enth}$
in \Tab{Summary}, we see that only $\beta$ enters, so the scaling of
$\urms$ is fully characterized by that of small blobs with negative buoyancy.

\section{Numerical solutions}
\label{NumSol}

In this section we present numerical solutions to demonstrate the effect
of the $\nabla_{\rm D}$ term on the resulting stratification.
At the end of this section, we also compare with the non-convecting
reference solution mentioned in the introduction.
We should emphasize that, although we use solar parameters, our models
cannot represent the Sun, because ionization effects have been ignored
(we take $\mu=0.6$ for the mean molecular weight in
$\cp-\cv={\cal R}/\mu$, where ${\cal R}$ is the universal gas constant).
A rather simple opacity law of the form of \Eq{doubleKramers} with
$\kappa_0=10^4\cm^2\g^{-1}$, $\rho_0=10^{-5}\g\cm^{-3}$, $T_0=13,000\K$,
and $a=0.5$, $b=18$, for the exponents in the power-law expression for
$\kappa_{\Hminus}$ has been used.
We also neglect the departure from plane-parallel geometry, so our
model can only give qualitative indications.

The system of two differential equations \eq{dlnP} and \eq{dlnT} decouples
by using $\ln P$ as the independent variable.
We thus integrate
\EQ
\dd\ln \meanT/\dd\ln\meanP=\nabla,
\label{dlnT_final}
\EN
using Equations \eq{nabD_ansatz}, \eq{fzeta}, and \eq{nabla_final}
to compute $\nabla$.
As an initial condition we use $\meanT_{\rm top}=\Teff/2^{1/4}$
at a sufficiently low pressure (here $\meanP_{\rm top}=10^5\dyn\cm^{-2}$),
so as to capture the initially isothermal part of the atmosphere; see,
e.g., \cite{Vit58} or \cite{Mih78}.
Here, $\Teff=(F_{\rm tot}/\sigmaSB)^{1/4}$ is the effective temperature.
We approximate the $\urms$ term in \Eq{urms} by using the value from the
previous step.
The Deardorff term is characterized by the assumed value of $f_{s0}$
and the value of $\tilde\zeta=\zeta-\Delta\zeta$ (\Sec{DepthDependence}),
where $\zeta=1$ and $\Delta\zeta$ is a function of $\beta$ and $\tbeta$
(\Sec{SuperadiabaticGradient}).
Since we integrate from the top downward, no prior knowledge of
$\rho_\ast$ is needed, because the Deardorff term is invoked only
after $\nabla-\nabad$ has reached its peak value.

Geometrical and optical depths are obtained respectively as
\EQ
-z=\int(\meanP/\meanrho g)\,\dd\ln \meanP
\quad\mbox{and}\;\;
\tau=\int(\kappa \meanP/g)\,\dd\ln \meanP,
\EN
where the integration of $\tau$ starts at $\ln\meanP_{\rm top}$ and that
of $-z$ at the position where $\tau=1$, which is referred to as the surface.
The factor $\kappa \meanP/g$, which is the same as $\HP$, is retained
because of the similarity with that in the expression for $\tau$.
The $z$ coordinate is used in some of our plots.

\begin{figure}[t!]\begin{center}
\includegraphics[width=\columnwidth]{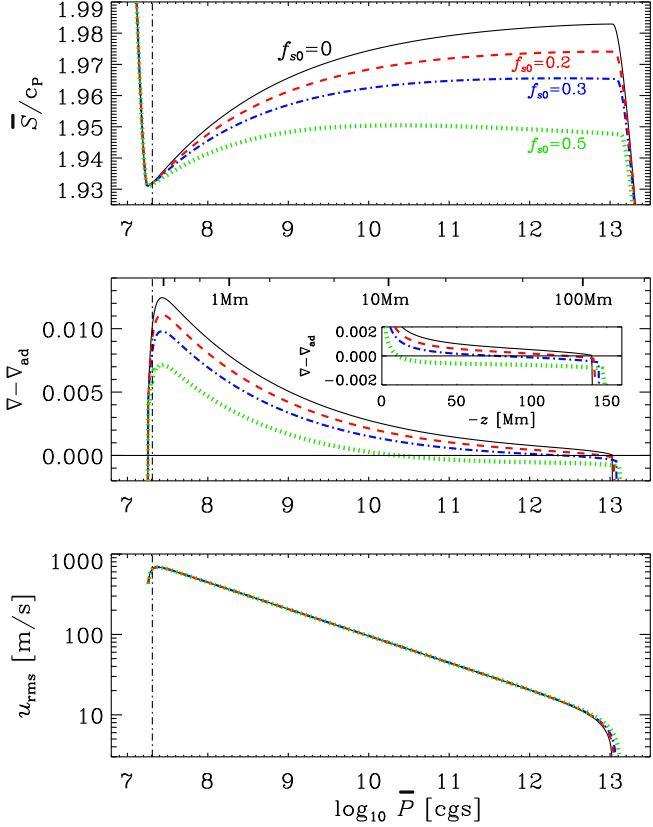}
\end{center}\caption[]{
Profiles of $\meanS/\cP$, $\nabla-\nabad$, and $\urms$ for $f_{s0}=0$
($\nabD=0$), as well as $f_{s0}=0.2$, $0.3$, and $0.5$ for
$\beta=\tbeta=0$ (Case~I) with $\tilde\zeta=1/9$.
The location of the surface ($\tau=1$) is indicated by vertical
dash-dotted lines and geometric depths below the surface are indicated
in the middle panel, starting with tick marks at $100$, $200$, and
$500\km$, and continuing with $1$, $2$, and $5\Mm$, etc.
The inset in the middle panel shows $\nabla-\nabad$ over a narrower
range as a function of $-z$.
}\label{pcomp5_lnppp_kfHp00}\end{figure}

\begin{figure}[t!]\begin{center}
\includegraphics[width=\columnwidth]{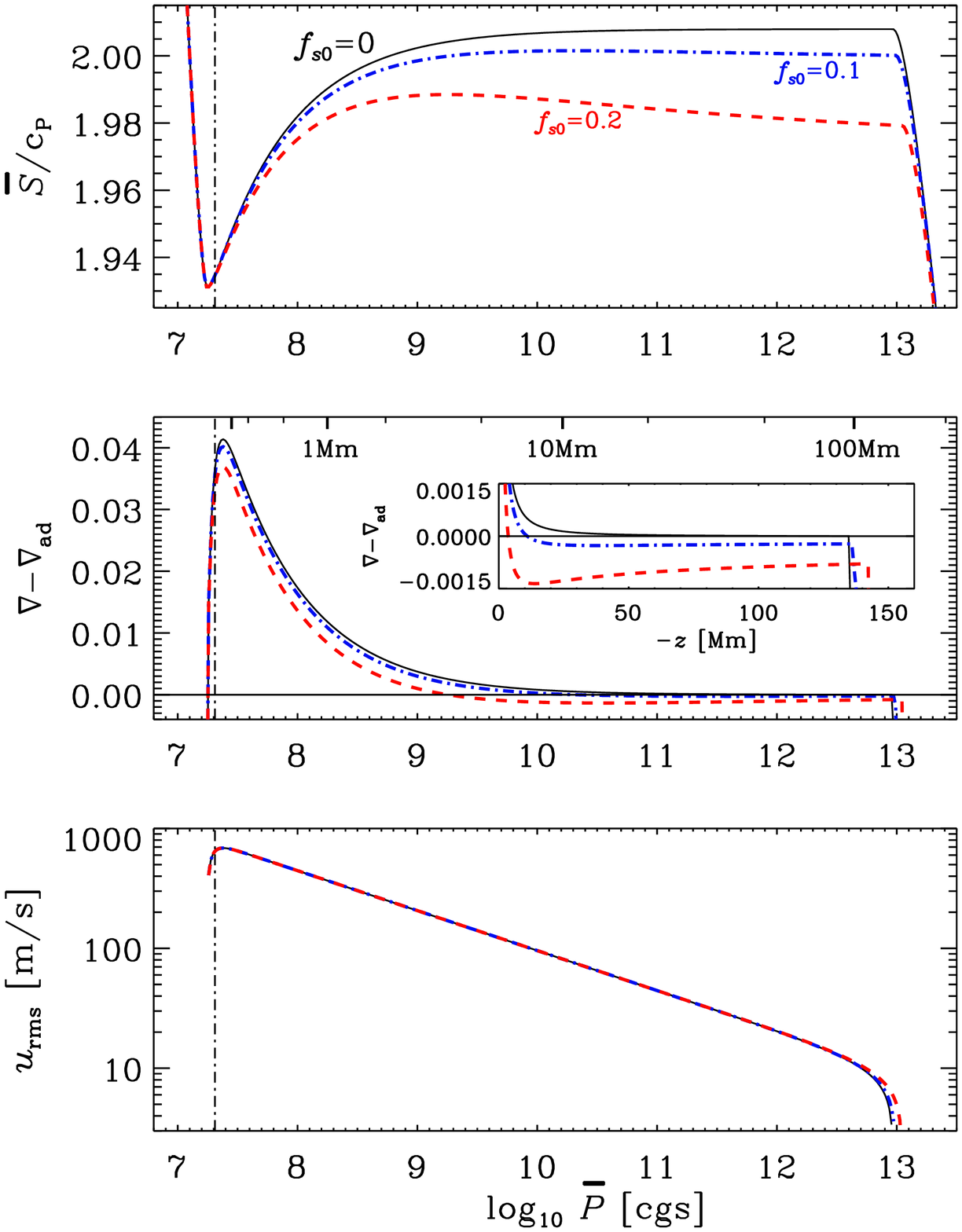}
\end{center}\caption[]{
Same as \Fig{pcomp5_lnppp_kfHp00}, but for $\beta=0$, $\tbeta=1$
(Case~II) with $\tilde\zeta=1/9$, and $f_{s0}=0$, $0.1$, and $0.2$.
}\label{pcomp5_lnppp_kfHp01}\end{figure}

In the following, we present solutions for the three combinations
of $\beta$ and $\tbeta$ sketched in \Fig{pbetacases2}.
We recall that only Cases~I and II ($\beta=0$ with $\tbeta=0$ or $1$)
correspond to small length scales in the deeper layers
(see also the power spectra in \Fig{pbetacases2}) and are therefore
of interest when trying to reconcile the non-detection of convective
motions by \cite{Hanasoge,HGS16} at the theoretically expected levels.
We did already emphasize that Case~II with $\beta=0$ and $\tbeta=1$ is
likely to lead to large kinetic energy fluxes.
However, based on the results presented below, it turns out that in the case
$\beta=\tbeta=0$ (Case~I), which was favored by these two requirements (small
length scale and non-divergent kinetic energy flux), the Deardorff term is
unlikely to have a significant effect, because for $\tbeta=0$, the gradient
term in the enthalpy flux becomes rather inefficient and must therefore
be compensated for by a correspondingly larger superadiabatic gradient,
and thus, only a rather large Deardorff flux ($f_{s0}\ge0.5$)
can make the resulting stratification sufficiently subadiabatic.
This is the case shown in \Fig{pcomp5_lnppp_kfHp00}, where we present
profiles of $\meanS/\cP$, $\nabla-\nabad$, and $\urms$ for $f_{s0}=0$
(no Deardorff term), as well as $f_{s0}=0.2$, $0.3$, and $0.5$.
The $\urms$ profiles are basically the same for all values of $f_{s0}$
and fall off like $\meanP^{\,-1/3}$.
This is expected, because $\urms\propto\meanT^{1/2-m}$ and $m=4/3$;
see \Eq{mexpression}, where we have used $\cs\propto\meanT^{1/2}$.
Thus, for isentropic stratification we have
$\urms\propto\meanP^{(1-2m)/5}=\meanP^{\,-1/3}$.

Next, we show in \Figs{pcomp5_lnppp_kfHp01}{pcomp5_lnppp_kfHp} Cases~II
and III with $\tbeta=1$ and $f_{s0}=0$, $0.1$, and $0.2$.
For both $\beta=0$ and $\beta=1$, the Deardorff flux now has a stronger
effect and is able to make the deeper parts of the domain Schwarzschild
stable even for $f_{s0}=0.1$.
Those layers would then be Schwarzschild stable and no longer a source
of giant cells.
Again, the profiles of $\urms$ are similar regardless of the
value of $f_{s0}$, but fall off more slowly for $\beta=1$
($\urms\propto\meanP^{\,-1/5}$), compared to $\beta=0$
($\urms\propto\meanP^{\,-1/3}$).
This agrees with our theory, because for $m=1$ we have
$\urms\propto\meanP^{(1-2m)/5}=\meanP^{\,-1/5}$.

Given that the stratification is stable in the deeper parts, we can
calculate the Brunt-V\"ais\"al\"a frequency of buoyancy oscillations,
$N_{\rm BV}$, which is given by $N_{\rm BV}^2=-(\nabla-\nabad)g/\HP$.
At intermediate and larger depths of the convection zone, we have
$g/\HP\la10^{-2}\s^{-2}$ and $\nabad-\nabla\la10^{-4}$, so the
period of buoyancy oscillations would be of the order of days.
This is comparable to or less than the turnover time $\tau$.
Indeed, one finds that
$\tau N_{\rm BV}\approx\sqrt{\nabD/\Delta\nabla-1}$
exceeds unity in the deeper parts, which is a consequence of $\nabD$
falling off with $\meanrho$ with a smaller power than $\Delta\nabla$,
as discussed in \Sec{RelativeImportance}.
Nevertheless, the resulting decrease in $\meanS/\cP$ with
depth remains always small ($10^{-3}$ or less) compared
with the value of $\Delta S_0/\cP$ produced at the surface
($\approx\urms^2/a_{\rm MLT}\nabad\cs^2\approx0.01$).
Thus, based on this, the descending low-entropy blobs would reach
the bottom of the convection zone before their negative buoyancy is
neutralized by the decreasing average entropy.
In other words, they will never perform any actual oscillations
before reaching the bottom of the convection zone,
i.e.\ where $F_{\rm conv}=0$.

\begin{figure}[t!]\begin{center}
\includegraphics[width=\columnwidth]{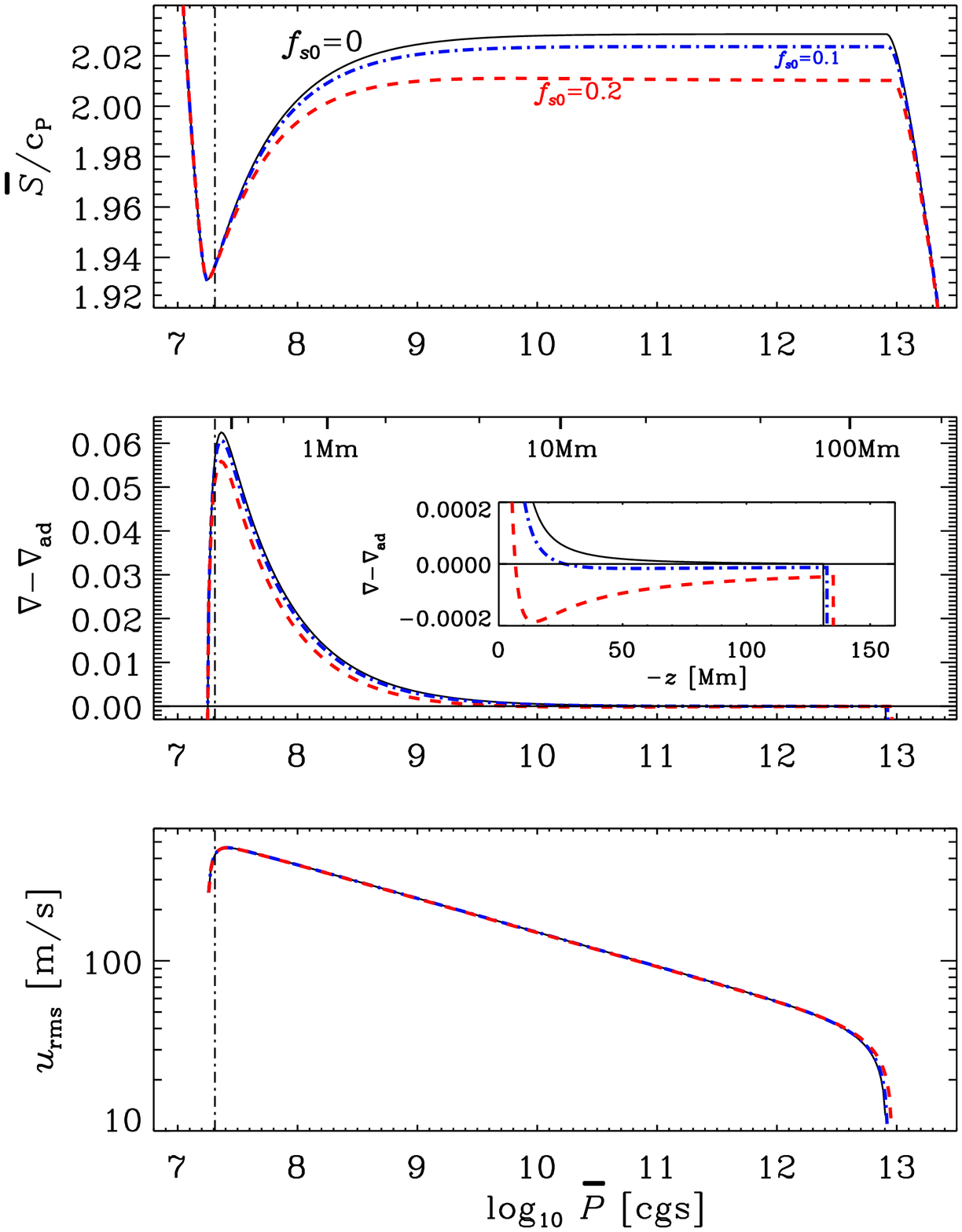}
\end{center}\caption[]{
Same as \Fig{pcomp5_lnppp_kfHp00}, but for $\beta=\tbeta=1$ (Case~III)
$\tilde\zeta=1/3$, and $f_{s0}=0$, $0.1$, and $0.2$.
}\label{pcomp5_lnppp_kfHp}\end{figure}

It may be interesting to note that the depth of the convection zone
increases slightly with increasing values of $f_{s0}$.
Of course, our model is idealized and represents the Sun at best
only approximately.
Furthermore, as emphasized in \Sec{Intro}, the depth of the convection zone
is well determined seismically, and this should be reproduced by a solar
model with realistic atomic physics and appropriately chosen
adjustable parameters.
However, it is known from the work of \cite{Ser09} that a slight expansion
of the solar convection zone would actually be required to compensate for
the shrinking that follows from the downward revision of solar
abundances \citep{Asp04} which is based on three-dimensional
convective atmosphere simulations, compared to previous analysis
based on one-dimensional semi-empirical models \citep{GS98}.

\begin{figure}[t!]\begin{center}
\includegraphics[width=\columnwidth]{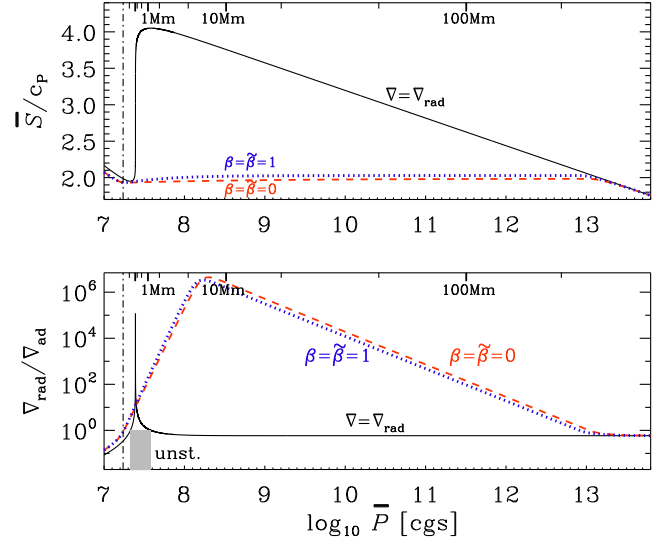}
\end{center}\caption[]{
Comparison of $\meanS/\cP$ (top) and $\nabrad/\nabad$ (bottom)
between the non-convective radiative reference solution ($\nabla=\nabrad$)
and standard convective solutions ($\nabla_{\rm D}=0$) with
$\beta=\tbeta=0$ (red, dashed) and $\beta=\tbeta=1$ (blue, dotted).
In both panels, the location of the $\tau=1$ surface is indicated by
vertical dash-dotted lines and geometric depths below the surface are
indicated for the non-convective solution, starting with tick marks at
$50$, $100$, and $200\km$, etc.
The location of the a priori unstable layer ($\nabrad>\nabad$) is marked
in the lower panel by a small gray strip.
For the other solutions, the depths are different; see those
in \Fig{pcomp5_lnppp_kfHp} for the solution with $f_{s0}=0$, which is
the same as here for $\beta=\tbeta=1$.
}\label{pcomp5_rad_ppp}\end{figure}

Finally, we compare in \Fig{pcomp5_rad_ppp} the standard convective
solution ($\beta=\tbeta=1$; same as the case with $f_{s0}=0$ in
\Fig{pcomp5_lnppp_kfHp}) with the non-convective radiative reference
solution.
Not surprisingly, owing to the absence of convection, the same flux
can now only be transported with a greatly enhanced negative (unstable)
entropy gradient near the surface.
However, this layer is now extremely thin ($1.15\Mm$) and the peak in 
$\nabrad/\nabad$ is about $100\km$ below the $\tau=1$ surface.
Note also that its peak value ($\approx10^5$) is below that in the
presence of convection, where it reaches a maximum of $\approx4\times10^6$
at a depth of $\approx1\Mm$.
As shown in \Eq{Nu}, this value can be interpreted as the local Nusselt number.
Note also that the result for $\beta=\tbeta=0$ (when $\nabD$ is weak)
is rather similar to that for $\beta=\tbeta=1$; see the dashed and
dotted lines in \Fig{pcomp5_rad_ppp}.

Our calculations have demonstrated that for $\tbeta=1$, regardless of the value
of $\beta$, bulk mixing changes the non-convecting reference state to a
nearly isentropic one.
However, whether the mean entropy gradient is slightly stably or slightly
unstably stratified depends on the presence of the Deardorff flux.
In the model with $\tbeta=0$, however, bulk mixing is rather inefficient
and the stratification would be Schwarzschild unstable unless an
unrealistically large Deardorff flux is invoked.

At the end of the introduction, we discussed that the entropy rain itself
might create an unstable stratification.
Let us now return to this question with more detailed estimates.
This will be done in the following section.

\section{Alternative considerations}
\label{UnstablyStratified}

Assuming that the upflows are perfectly isentropic, \cite{Spr97} argues
that the low-entropy material from the top with its decreasing entropy
filling factor $f_s$ in deeper layers necessarily leads to a negative
mean entropy gradient.
Specifically, using \Eq{meanSfill}, one obtains
\EQ
-\HP{\dd\meanS/\cP\over\dd z}
={(\Delta S)_0\over\cP}{\dd f_s\over\dd\ln\meanP}
={2\over5}{(\Delta S)_0\over\cP}f_s > 0,
\EN
where we have used $\dd\ln f_s/\dd\ln\meanrho=2/3$
and $\dd\ln\meanrho/\dd\ln\meanP=3/5$ for an isentropic
layer with $\gamma=5/3$.
Thus, the stratification would be Schwarzschild unstable.
This is also borne out by the solar simulations with realistic
physics, although the computational domains are sufficiently
shallow so that the radiative flux is still small in the deeper
parts and usually even neglected altogether.
Toward the bottom of the convection zone, however, radiation becomes
progressively more important and the mean entropy gradient in the upflows
may no longer be vanishing.

To estimate the mean entropy gradient in the upflows, we may balance
the steady state entropy advection with the negative radiative flux
divergence, i.e.
\EQ
\meanrho\,\meanT\,\meanU_{\uparrow}\,\dd\meanS_\uparrow/\dd z
\approx-(\dd F_{\rm rad}/\dd z)_\uparrow.
\label{dSdz_up}
\EN
The sign of $\dd F_{\rm rad}/\dd z$ is negative and thus compatible
with a positive $\dd\meanS_\uparrow/\dd z$ in the upflows, but it
would only be large enough near the bottom of the convection zone.
Higher up, $\dd F_{\rm rad}/\dd z$ becomes smaller and eventually unimportant.
On the other hand, $\meanU_\uparrow/\urms$ can be rather small
(see \Tab{SummaryPHIKIN}).
Furthermore, our considerations neglect the fact that the gas in the upflows
expands, so only a fraction of the gas can ascend before it begins to
occupy the available surface area.
Therefore, the rest of the gas would have to remain stagnant
and continue to heat up.
In reality, of course, there would be continuous entrainment, resulting
in a finite, but still low, effective upward velocity.

\begin{figure}[t!]\begin{center}
\includegraphics[width=\columnwidth]{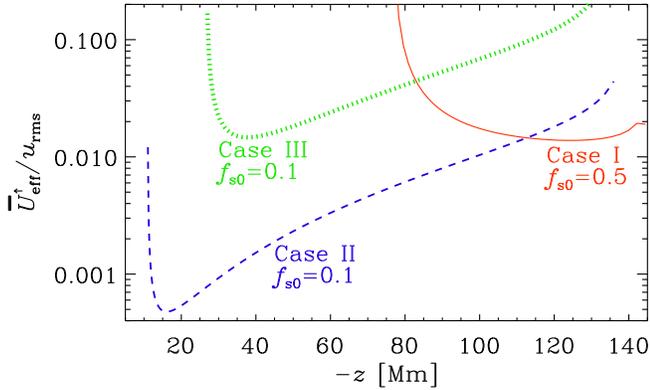}
\end{center}\caption[]{
Dependence of $\meanU^\uparrow_{\rm eff}/\urms$ on depths $-z$
for Case~I with $f_{s0}=0.5$ (solid red line), as well as Cases~II and III
with $f_{s0}=0.1$ (dashed blue and dotted green lines, respectively).
}\label{p2}\end{figure}

We now use selected solutions obtained in \Sec{NumSol} to estimate
the effective fractional upward velocity for radiative heating/cooling
to dominate over advection, defined as
\EQ
\meanU^\uparrow_{\rm eff}/\urms=
(-\dd F_{\rm rad}/\dd z)\,/\,(\urms\,\meanrho\,\meanT\dd\meanS/\dd z).
\EN
Here we have used $\meanT\dd\meanS/\dd z=g\,\Delta\nabla/\nabad$.
\Fig{p2} shows this quantity for Cases~I--III, which are shown in
\Figss{pcomp5_lnppp_kfHp00}{pcomp5_lnppp_kfHp}.
It turns out that for Case~I with $f_{s0}=0.5$, the effective fractional
upward velocity is around 0.01, while for Case~III with $f_{s0}=0.1$
it can be even larger.
The value 0.01 is compatible with the $\meanU^\uparrow\!/\urms$
values in \Tab{SummaryPHIKIN} if we assume $f=0.01$
(so $\meanU^\uparrow\!/\urms=0.1$) and
$\meanU^\uparrow_{\rm eff}/\meanU^\uparrow=0.1$.
For Case~II with $f_{s0}=0.1$, the effective fractional upward velocity is
around $10^{-3}$, which may be unrealistically small.

Based on the considerations above, we may conclude that the case for a
Schwarzschild-stable entropy gradient depends on model assumptions for
the implementation of the Deardorff flux that are potentially in conflict
with the mean entropy gradient expected from the two-stream model.
Whether or not there is really a potential conflict depends not only on
the model parameters, but also on \Eq{dSdz_up} itself.
This is because the $\meanT$ factor in the entropy equation is outside
the derivative, making it impossible to derive the total flux balance
assumed in \Eq{FluxBalance}; see also \cite{RK15} for related details.

In view of these complications, it is worthwhile to discuss alternative
ways of avoiding giant cell convection.
For this purpose, we have to address the global problem of using
some coarse-grained form of effective mean-field equations.
When considering the equations of mean-field hydrodynamics, in which the
small-scale enthalpy and momentum fluxes are parameterized in terms
of negative mean entropy and mean velocity gradients, respectively,
one finds solar differential rotation as a result of non-diffusive
contributions to the Reynolds stress, but in certain parameter regimes
a new instability was found to develop \citep{TR89,R89,RS92}.
This instability was later identified as one that is analogous to
Rayleigh-B\'enard convection, but now for an already convecting mean
state \citep{TBMR94}.
This instability would lead to giant cell convection.

The existence of giant cell convection is under debate, but assuming that it
does {\em not} exist in the Sun, one might either hypothesize that the
mean-field equations are too simplified or that mean-field convection
could be suppressed by sufficiently strong turbulent viscosity and
turbulent thermal diffusivity coefficients.
These turbulent coefficients define a turbulent Rayleigh number for a
layer of thickness $d$,
\EQ
\Rat={gd^4\over\nut\chit}\,
\left({\Delta\nabla\over \HP}\right)_{z_*}.
\label{Raturb}
\EN
This number would then have to be still below the critical value for
convection.
\EEq{Raturb} differs from the usual one defined through \Eq{RaDef} in
\App{Rayleigh} in that, first, $\nu\to\nut$ and $\chi\to\chit$ have been
substituted, and second, the superadiabatic gradient of the
non-convecting reference solution is now replaced by the actual one.
This might be a plausible alternative to explaining the absence of
giant cell convection if the idea of turning the stratification from
Schwarzschild unstable to Schwarzschild stable through the $\nabD$ term
were to turn out untenable.

We recapitulate that in this alternate explanation, the stratification
is Schwarzschild unstable, i.e.\ $\Rat>0$, corresponding to \Eq{unstable},
but $\Rat<\Ratcrit$ is still below a certain critical value, so it would
be stable by a turbulent version of the Rayleigh-B\'enard criterion.
\cite{TBMR94} found $\Ratcrit\approx300$ for a vanishing rotation rate.
However, they also estimated that $\Rat>\Ratcrit$ for plausible solar
parameters, so the direct adoption of this idea would be problematic, too.
However, one might speculate that a more accurate treatment could lead
to stability when allowing, for example, for spatial nonlocality of the
turbulent transport \citep{BRS08,RB12}.

\section{Conclusions}
\label{Conclusions}

In the present work we have suggested that the enthalpy flux in stellar
mixing length models should contain an extra nonlocal contribution so
that the enthalpy flux is no longer proportional to the local
superadiabatic gradient, $\nabla-\nabad$, but to $\nabla-\nabad+\nabD$,
where $\nabD$ is a new nonlocal contribution that was first identified
by \cite{Dea66} in the meteorological context.
The significance of this term lies in the fact that it provides an
alternative to the usual local entropy gradient term and can transport
enthalpy flux outwards---even in a slightly stably stratified layer.

We have presented a modified formulation of stellar MLT that includes
the $\nabD$ term, in addition to a $\nabkin$ term resulting from the
kinetic energy flux.
The formalism and the final results are similar to those of conventional MLT
in that one also arrives here at a cubic equation for $\nabla$, but the term
$\nabla-\nabad$ is now replaced by $\nabla-\nabad+\nabD-\nabkin$.
This new formulation implies that convection can carry a finite flux while
$\nabla-\nabad$ is still negative and therefore the stratification
is Schwarzschild stable, i.e., \Eq{stable} is obeyed.
Consequently, if confirmed, no large length scales are being excited.

The present formulation allows for different treatments of the length scales
governing buoyant elements on the one hand and the time and length scales
associated with mixing on the other.
When both are independent of depth ($\beta=\tbeta=0$), mixing becomes
inefficient at larger depths.
Thus, to carry a certain fraction of the
enthalpy flux, the superadiabatic gradient needs to be larger than
otherwise, making it harder for the Deardorff term to revert the sign
of $\nabla-\nabad$.
On the other hand, for a tree-like hierarchy of many downdrafts merging
into fewer thin ones at greater depths ($\beta=0$, $\tbeta=1$), the
increasing length scale associated with increasing
separation enhances vertical mixing, making the stratification
nearly isentropic without the Deardorff term, and slightly subadiabatic
with a weak Deardorff term.
In that case, however, the filling factor of the downflows decreases
with depth as $\rho^{-\zeta}$, which may imply an unrealistically
large downward kinetic energy flux.
This leaves us with the standard flow topology ($\beta=\tbeta=1$), where
both size and separation of structures increase with depth.
The Deardorff flux can still cause the stratification to have a
subadiabatic gradient, so no giant cell convection would be excited
locally, but the flow structures would be large and should be
helioseismically detectable, as has been found by \cite{Greer} using
ring-diagram local helioseismology.

It would be useful to explore the thermodynamic aspects of the present
model more thoroughly and to connect with related approaches.
An example is the work by \cite{Rem04}, who studied a semianalytic
overshoot model that was driven nonlocally by downdraft plumes,
similar to what was suggested by \cite{Spr97}.
Rempel also finds an extended subadiabatic layer in large parts
of the model.
Furthermore, there are similarities to the nonlocal mixing length model by
\cite{XD01}, who, again, find an extended deeper layer that is subadiabatically
stratified in the deeper parts of their model.
In this connection we emphasize the main difference between nonlocal
turbulence owing to the Deardorff flux and usual overshoot:
in the latter case the enthalpy flux would go inward
as a consequence of the reversed entropy gradient,
while in the present model the Deardorff flux goes outward.

Future work could proceed along two separate paths.
On the one hand, one must establish the detailed physics leading to the
$\nabD$ term using models with reduced opacity, in which reliable DNS
are still possible, i.e., no SGS terms are added and the primitive
equations are solved as stated, without invoking \Eq{FradSplit}.
On the other hand, one could study suitably parameterized large eddy
simulations that either include a nonlocal Deardorff term of the form
given by \Eq{nabD_ansatz}, as discussed in \Sec{Deardorff},
or that explicitly release entropy rain at the surface
such that the resulting stratification is still slightly stable.
This would be particularly useful in global simulations that
would otherwise have no entropy rain.

As stimulating as the results of \cite{Hanasoge} are, they do require
further scrutiny and call for the resolution of the existing conflicts
with other helioseismic studies such as those of \cite{Greer}.
Alternatively, global helioseismic techniques for detecting giant cell
convection \citep{LR93,CA09,Woo14} can provide another independent way
of detecting deep larger-scale flows \citep{Woo16}.
Realistic simulations should eventually agree with helioseismic
results of flows in deeper layers of the Sun, but at the moment it is
still unclear whether a subgrid scale treatment as in
\Eq{FradSplit} adequately captures
the small-scale flows that can be responsible for the Deardorff flux and
whether they would in principle be able to predict the subtle departures
from superadiabatic stratification on subthermal time scales.

\acknowledgments
I thank the referee for his/her criticism that has led to many
improvements, and Evan Anders, Jean-Francois Cossette, Ben Greer,
{\AA}ke Nordlund, Mark Rast, Matthias Rempel, Matthias Rheinhardt,
Bob Stein, Peter Sullivan, Regner Trampedach, and J\"orn Warnecke
for interesting discussions and comments.
I also wish to acknowledge Juri Toomre for mentioning to me the need for
modeling entropy rain during my time in Boulder some 24 years ago.
This work was supported in part by
the Swedish Research Council grant No.\ 2012-5797,
and the Research Council of Norway under the FRINATEK grant 231444.
This work utilized the Janus supercomputer, which is supported by the
National Science Foundation (award number CNS-0821794), the University
of Colorado Boulder, the University of Colorado Denver, and the National
Center for Atmospheric Research. The Janus supercomputer is operated by
the University of Colorado Boulder.

\appendix

\section{Polytropic stratification from Kramers opacity}
\label{Kconst}

We show here that, for the non-convecting reference solution, using the
Kramers-like opacity law of \Eq{Kramers}, but not the combined opacity
law of \Eq{doubleKramers}, we have $K\to\const$ in the deeper, optically
thick layers.
Dividing \Eq{FradEqn} by the equation for hydrostatic equilibrium,
$\dd P/\dd z=-\rho g$, we have
\EQ
{\dd T\over\dd P}
={F_{\rm rad}\over K\rho g}
={F_{\rm rad}\over K_0\rho_0 g} {(\rho/\rho_0)^a\over (T/T_0)^{3-b}}
={F_{\rm rad}\over K_0\rho_0 g} {(P/P_0)^a\over (T/T_0)^{3+a-b}},
\label{dTdP}
\EN
where $K_0=16\sigmaSB T_0^3/(3\kappa_0\rho_0)$ is a constant
and $P/P_0=(\rho/\rho_0)(T/T_0)$ is the ideal gas equation with a
suitably defined constant $P_0=(\cp-\cv)\rho_0 T_0$.
Here, $\rho_0$ and $T_0$ are reference values that were
defined in \Eq{Kramers}.
\EEq{dTdP} can be integrated to give
\EQ
(T/T_0)^{4+a-b}=(n+1)\nabrad^{(0)}(P/P_0)^{1+a}+(T_{\rm top}/T_0)^{4+a-b},
\label{ToverT0}
\EN
where $\nabrad^{(0)}=F_{\rm rad}P_0/(K_0T_0\rho_0 g)$, which is
defined analogously to the $\nabrad$ without superscript $(0)$ in
\Eqs{nabla}{nabrad_Def}, and $T_{\rm top}$ is an integration constant
that is specified such that $T\to T_{\rm top}$ as $P\to0$.
Note also that $4+a-b=(n+1)(1+a)$, where $n$ was defined in \Eq{n_Def}
as the polytropic index, so the ratio of $4+a-b$ to $1+a$ is just $n+1$,
which enters in front of the $\nabrad^{(0)}$ term in \Eq{ToverT0}.
Since $K\propto T^{3-b}/\rho^{1+a}\propto T^{4+a-b}/P^{1+a}$,
we have $K\to\const=K_0$ for $T\gg T_{\rm top}$.

\section{Derivation of Equations (10) and (11)}
\label{deriv10and11}

To obtain \Eqs{sdot}{udot}, which are used to derive the Deardorff flux term in
the $\tau$ approximation, we start with the equations for specific entropy
and velocity in the form \citep[see, e.g.,][]{BB14}
\EQA
\label{dss}
\rho T{\DD S \over \DD t}&=&-\nab\cdot\FF_{\rm rad},\\
\label{duu}
\rho{\DD \UU\over \DD t}&=&-\nab P +\rho\grav,
\ENA
where $\DD / \DD t= \partial /{\partial t} + \UU \cdot \nab$ is the
advective derivative, $\FF_{\rm rad}$ is the radiative flux,
and viscosity has been omitted.
Subtracting those equations from their averaged ones,
we obtain the following set of equations
\EQA
\label{dssf}
{\partial s \over \partial t}
+\meanU_j{\partial s \over \partial x_j}
+u_j{\partial \meanS \over \partial x_j} 
&=&-{1\over\meanrho\,\meanT}\nab\cdot\delta\FF_{\rm rad}+{\cal N}_s,\\
\label{duuf}
{\partial u_i \over \partial t}
+\meanU_j{\partial u_i \over \partial x_j}
+u_j{\partial \meanU_i \over \partial x_j} 
&=&-{1\over\meanrho}\nab p+g_i(p/\gamma\meanP-s/\cP)+{\cal N}_{ui},
\ENA
where we have used $\delta\rho/\meanrho=p/\gamma\meanP-s/\cP$,
which can be obtained from \Eq{Seqn} and the perfect gas law
by linearization.
Pressure fluctuations will again be neglected and
$(\nab\cdot\delta\FF_{\rm rad})/\meanrho\,\meanT$ will be replaced
by $-s/\tau_{\rm cool}$, as explained in \Sec{Deardorff}.
Assuming $\meanUU=\bm{0}$ and omitting the nonlinear terms ${\cal N}_s$
and ${\cal N}_{ui}$ we arrive at \Eqs{sdot}{udot}.

\section{Filling factor for a descending Hill vortex}
\label{HillVortices}

The solution for a Hill vortex with radius $a_{\rm H}$ and propagation
velocity $u_{\rm H}$ is given by a stream function $\Psi$ in spherical
coordinates $(r,\theta,\phi)$ as \citep[e.g.][]{MM78}
\EQ
\Psi={1\over4}\left\{
\begin{array}{ll}
-3u_{\rm H}(1-r^2/a_{\rm H}^2)\,\varpi & \mbox{(for $r<a_{\rm H}$)},\\
+2u_{\rm H}(1-a_{\rm H}^3/r^3)\,\varpi & \mbox{(for $r>a_{\rm H}$)}.
\end{array}
\right.
\EN
We apply it in Cartesian coordinates as the initial condition for the
mass flux as $\rho\uu=\nab\times(\Psi\pphi)$, where
$\pphi=(-y/\varpi, x/\varpi, 0)$ is the unit vector in the toroidal
direction of the vortex, using $\varpi^2=x^2+y^2$ and $r^2=\varpi^2+z^2$
for cylindrical radius $\varpi$ and spherical radius $r$.
We adopt here an isothermal equation of state, i.e., there is no
buoyancy force in this problem.
For non-isothermal calculations, but in two dimensions, we refer to the
work of \cite{Ras98}.
Density and pressure fall off exponentially with height and both the
scale height and the sound speed are independent of height.
No analytic solution exists in that case, so the Hill vortex solution
is at best approximate.
We consider a domain of size $L\times L\times 4L$ with
$-L/2<x,y/\HP<L/2$ and $-7L/2<z<L/2$, where $L=5\HP$.
We choose $a_{\rm H}=0.5\HP$ and $u_{\rm H}=0.2\cs$.
The viscosity is $\nu=5\times10^{-5}$, so the Reynolds number is
$a_{\rm H} u_{\rm H}/\nu=2000$.
We use the {\sc Pencil Code}\footnote{\url{https://github.com/pencil-code}}
with a resolution of $1152\times1152\times4608$ meshpoints.

\begin{figure*}[t!]\begin{center}
\includegraphics[width=.24\textwidth]{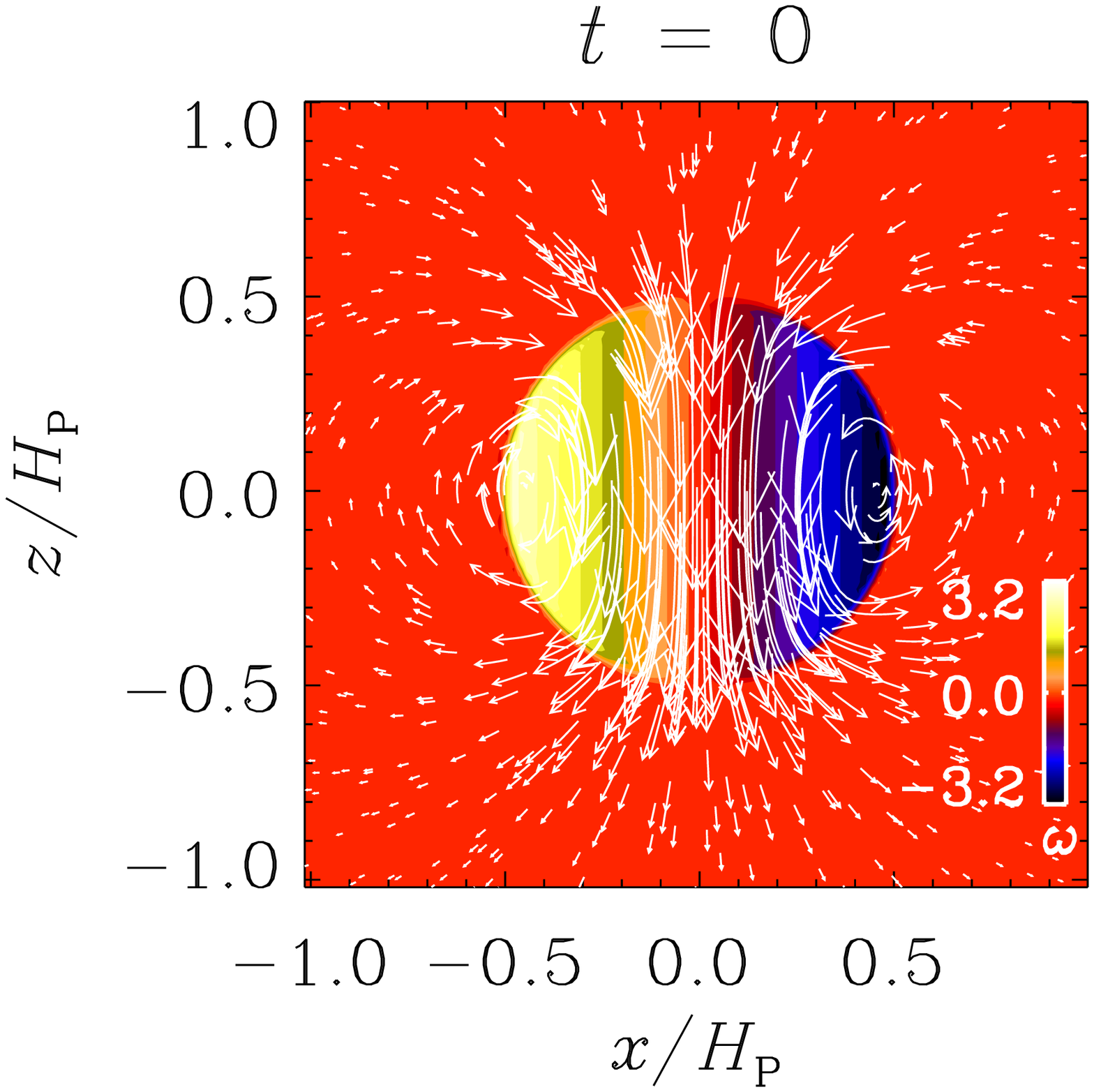}
\includegraphics[width=.24\textwidth]{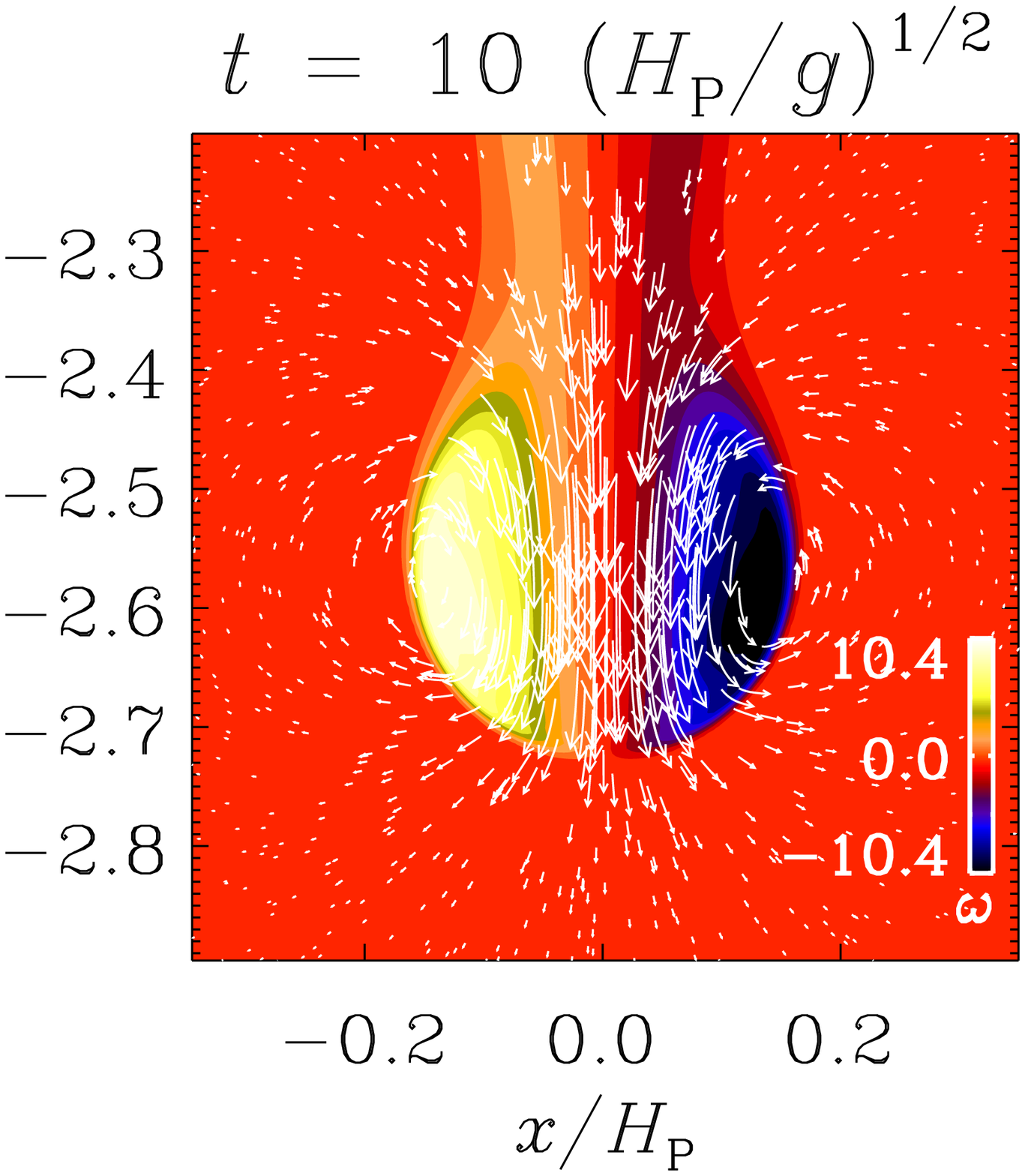}
\includegraphics[width=.24\textwidth]{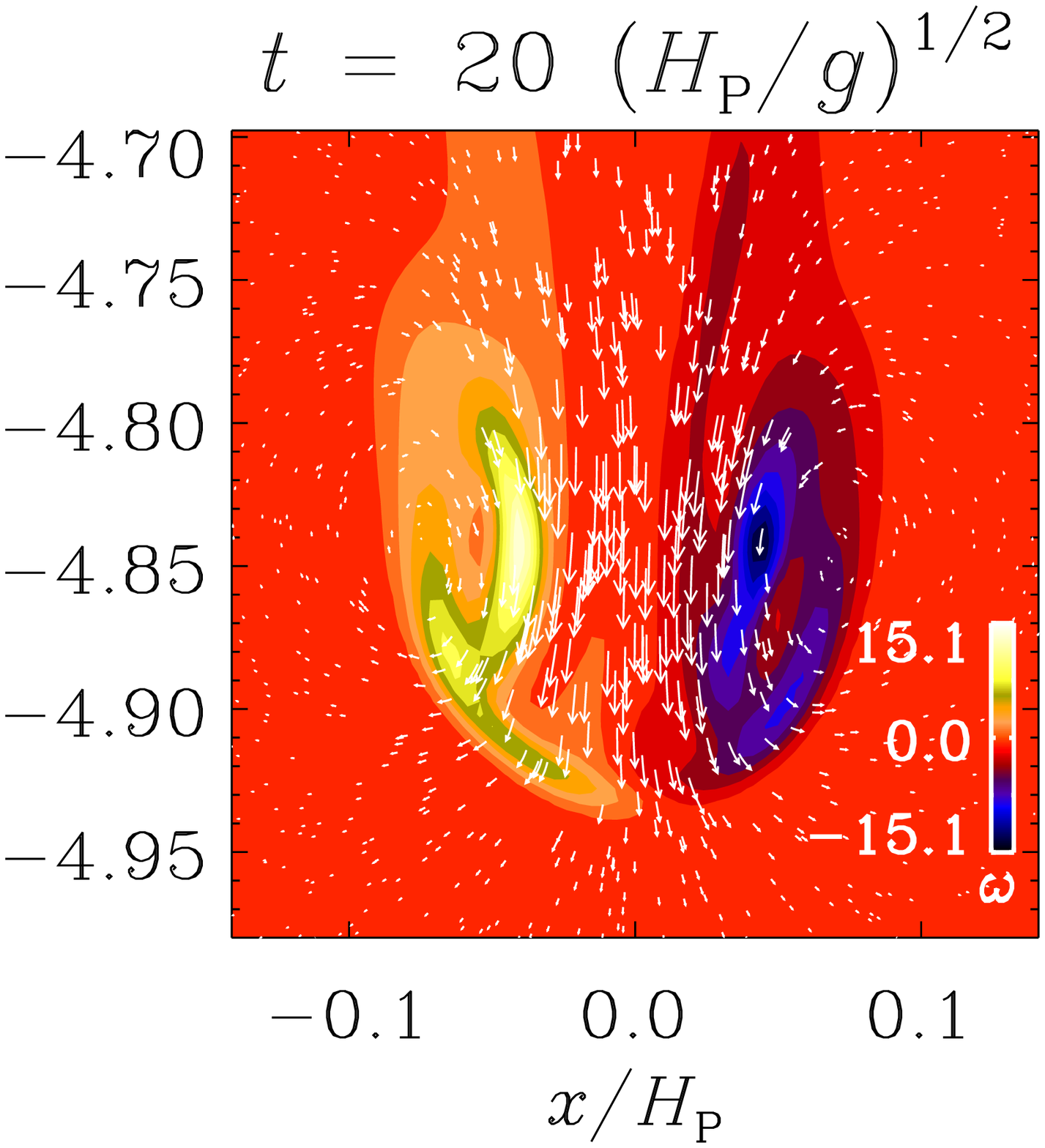}
\includegraphics[width=.24\textwidth]{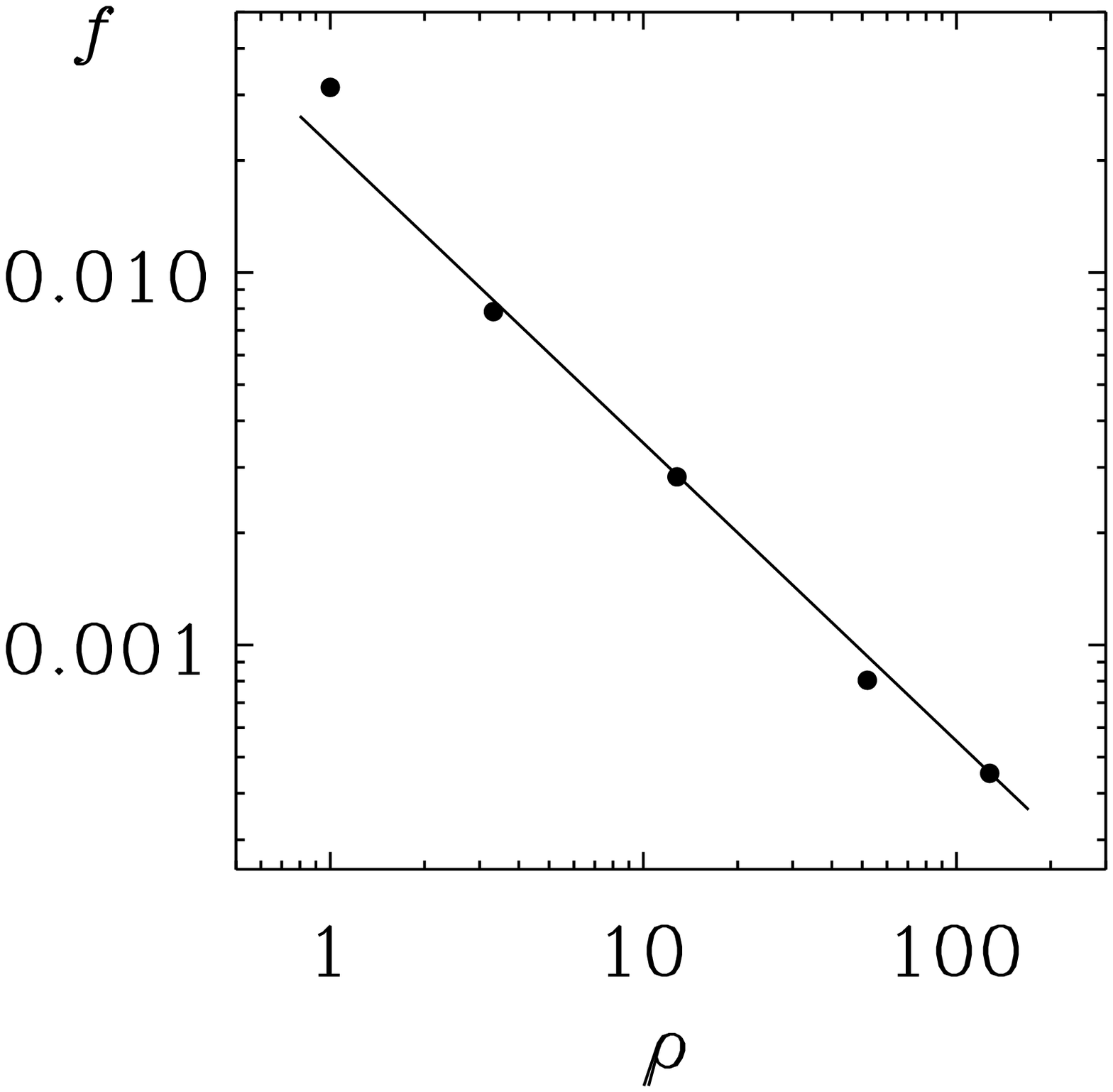}
\end{center}\caption[]{
Velocity vectors superimposed on a color scale representation of
the vorticity $\omega$ at times 0, 10, and 20
in units of $(\HP/g)^{1/2}$, as well as the resulting filling
factor versus density. The negative slope is $\zeta=0.8$.
Note that the frame of view changes as the vortex descends and shrinks.
}\label{rslice}\end{figure*}

\FFig{rslice} shows snapshots
zoomed into the vortex as it traverses about five scale heights.
The filling factor, which is proportional to the radius squared,
decreases with depth and is found to scale with the surrounding
density like \Eq{fsDef} with $\zeta=0.8$; see the last panel of
\Fig{rslice}.

\section{MLT relation between $\urms$ and $\srms$}
\label{DeltaS_Ma2}

To find the relation between $\urms$ and the rms values of temperature or
entropy fluctuations, it is customary in standard MLT to approximate the steady
state momentum equation, $\uu\cdot\nab\uu\approx-\grav\,\delta T/\meanT$,
by $\urms^2\kf=a_{\rm MLT}\,g\,\delta T/\meanT$, where $a_{\rm MLT}\approx1/8$
is a commonly adopted geometric factor \citep[e.g.,][]{Spr74}.
This leads to
\EQ
\srms/\cP=\delta T/T = \urms^2\kf/a_{\rm MLT}g
=\gamma \Ma^2\kf \HP/a_{\rm MLT},
\label{srms}
\EN
where we have used $\cs^2=\gamma g\HP$, and thus
$(\srms/\cp)\,\Ma^{-2}=\gamma\kf\HP/a_{\rm MLT}$.
Furthermore, using $\nabad\cp\meanT=g\HP$, we have
$\meanT\srms=\urms^2\kf \HP/(a_{\rm MLT}\nabad)$,
which is used to derive \Eq{Fenth2}.

\section{Estimate for the surface value of $\nabD$}
\label{SurfaceValue}

The purpose of this section is to show that $\nabD$ is a certain fraction
of $\nabla-\nabad$ in the top layers, as stated in \Eq{nabD_ansatz}.
To have an estimate for $\overline{s^2}$, we multiply
\Eq{sdot} by $s$ and average, so we get
\EQ
{1\over2}{\partial\overline{s^2}\over\partial t}=
-\overline{u_j s} \, \nabla_j\meanS
-\overline{s^2}/\tau_{\rm cool}
+{\cal T}_s.
\label{s2eqn0}
\EN
As for $F_{\rm enth}$ in \Eqs{dotFconv}{MTAclosure}, we have a triple
correlation term, which is here ${\cal T}_s=-\overline{s\uu\cdot\nab s}$.
Again, we adopt the $\tau$ approximation and replace ${\cal T}_s$ by
a damping term of the form ${\cal T}_s=-\overline{s^2}/\tau$ which,
together with the $\tau_{\rm cool}$ term, combines
to give $\taured$ as the relevant time scale.
Assuming a statistically steady state, $\partial\overline{s^2}/\partial t=0$,
we derive the following expression for $\overline{s^2}$:
\EQ
\overline{s^2}=-\taured \, \overline{u_j s} \, \nabla_j\meanS
\quad\mbox{(in the surface layers)}.
\label{s2eqn}
\EN
This shows that fluctuations of specific entropy are produced when
there is an outward flux ($\overline{u_z s}>0$) and a locally negative
(unstable) mean entropy gradient; see also \cite{Garaud10} for a
similar derivation.
Inserting this into \Eq{FD}, and using \Eq{Fenth}, we obtain
\EQ
\FF_{\rm D}=\taured^2 \, \grav \, (\FF_{\rm enth}\cdot\nab\meanS/\cP)
\quad\mbox{(in the surface layers)}.
\label{FDterm}
\EN
Here, both $\grav$ and $\nab\meanS$ point downward, so $\FF_{\rm D}$
points upward and we can write $F_{\rm D}=\lambda F_{\rm enth}$, where
$\lambda=(\taured/\tau_{\rm ff})^2(\nabla-\nabad)$ is a
coefficient that itself is proportional to the superadiabatic gradient,
and $\tau_{\rm ff}=(\HP/g)^{1/2}$ is the free-fall time (after which a fluid
parcel at rest has reached a depth of $\HP/2$).
Since $F_{\rm enth}=F_{\rm G}+F_{\rm D}$, this implies
$F_{\rm G}=(1-\lambda) F_{\rm enth}$ and therefore $\lambda<1$
in the highly unstable layer at the top, where both $F_{\rm G}$ and
$F_{\rm enth}$ are positive.
We expect $F_{\rm D}$ to be largest just a few hundred kilometers
below the photosphere,
so $\lambda$ should be maximum in the upper parts.
To calculate the fraction $\nabD/(\nabla-\nabad)$, we use the fact
that $F_{\rm enth}=F_{\rm G}+F_{\rm D}$, together with the part of
\Eq{Fconv_urms2} that relates to $F_{\rm D}$, to write
\EQ
F_{\rm D}=\onethird\meanrho \cP \meanT \, (\taured\urms^2/\HP) \nabD
=(\taured/\tau_{\rm ff})^2(\nabla-\nabad)F_{\rm enth}.
\label{Fconv_urms2b}
\EN
Starting with \Eq{Fenth2} and expressing $\taured$ in terms of $\sigma$
using \Eq{urms}, we find
\EQ
\nabD^{\max}/(\nabla-\nabad)_{\max}=3\sigma\phi_{\rm enth}\nabad\
\alpha_{\rm mix}^{-\tbeta} (\kfz\HP)^{-(1-\tbeta)}\approx3,
\EN
where we have used $\tbeta=1$ (appropriate to the near-surface layers),
ignored radiative cooling ($\sigma=1$), and assumed $\alpha_{\rm mix}=1.6$
\citep{Sti02}, as well as $\phi_{\rm enth}\approx4$ that was found in
simulations of \cite{BCNS05}, as discussed in \Sec{DepthDependence}.

We emphasize that these equations only characterize the {\em initiation}
of entropy rain.
They cannot be used to compute the Deardorff flux in the deeper
layers where we have instead invoked Spruit's concept of nonlocal
convection in the form of threads.
The $\overline{s^2}$ associated with those threads is
likely to result from a part of the triple correlation term
${\cal T}_s=-\overline{s\uu\cdot\nab s}$, which gives rise to 
a negative divergence of the flux of $\overline{s^2}$ of the form
$\HH\equiv\overline{\uu s^2}$ on the rhs of \Eq{s2eqn0}.
This flux (which should not be confused with an energy flux) should
point downward ($s^2$ is largest when $u_z<0$) and be strongest in the
upper layers, so $\nab\cdot\HH<0$, leading to a positive contribution from
$-\nab\cdot\HH$ on the rhs of \Eq{s2eqn0}.
This may explain the $\overline{s^2}$ associated with the $\nabD$ term.

To estimate the ratio $F_{\rm D}/F_{\rm G}$ in the deeper layers,
we use \Eq{s2eqn0} in the steady state with $1/\tau_{\rm cool}\to0$
and ${\cal T}_s=-\nab\cdot\HH\approx(2\zeta-\Delta\zeta)\overline{s^2}
\urms/\gamma\HP$ and obtain
\EQ
0=-\overline{u_j s} \, \nabla_j\meanS
+(2\zeta-\Delta\zeta)\overline{s^2}\urms/\gamma\HP
\quad\mbox{(in deeper layers)}.
\EN
Multiplying by $\onethird\taured\urms^2 \,(\meanrho\,\meanT)^2$,
using \Eqs{Fenth}{FG}, and expanding the fraction by $g/\cp$, we have
\EQ
0=F_{\rm enth}\,F_{\rm G}
+\onethird\meanrho\urms^3(2\zeta-\Delta\zeta)\left(\taured
\overline{s^2} g \,\meanrho\,\meanT\right) \, \cp\meanT/(\gamma g\HP).
\EN
This implies that $F_{\rm G}<0$ in the deeper layers.
The second term in the parentheses is $F_{\rm D}$ (with a plus sign,
because $g>0$ is a scalar here); see \Eq{FD}.
Furthermore, we use $\cp\meanT/(\gamma g\HP)=1/(\gamma-1)$ and
$\meanrho\urms^3=F_{\rm enth}/\phi_{\rm enth}$; see \Eq{Fenth2}.
The $F_{\rm enth}$ terms on both sides cancel, so we have
\EQ
F_{\rm D}/|F_{\rm G}|=3(\gamma-1)/\left[\phi_{\rm enth}
(2\zeta-\Delta\zeta)\right]\approx0.3\,...\,0.5,
\EN
where $2\zeta-\Delta\zeta=2\tilde\zeta+\Delta\zeta=10/9$
for Cases~I and II, and $14/9$ for Case~III (\Sec{SuperadiabaticGradient}),
and $\phi_{\rm enth}=4$ has been assumed.

\section{Derivation of expression for \MakeLowercase{\mbox{$c_0$}}}
\label{App}

To find the coefficient $c_0$ given by \Eq{urms0},
we use \Eq{srms} in the form
\EQ
\cP \delta T=\urms^2\kf \HP/\nabad a_{\rm MLT}.
\EN
Inserting this into \Eq{Fconv_orig}, and using \Eq{kfbeta}, yields
\EQ
F_{\rm enth}=\meanrho\urms^3\,\alpha_{\rm mix}^{\beta}
(\kfz \HP)^{1-\beta}/\nabad a_{\rm MLT}.
\EN
Equating this expression with \Eq{Fconv_urms1}, using \Eq{F0},
we can derive the desired expression for $\urms$ in the form
\EQ
\urms^2={\sigma\over3}{\cs^2\over\gamma}{a_{\rm MLT}\over(\kfz \HP)^{2-2\beta'}}
\left(\nabla-\nabad+\nabD\right),
\label{urms0_orig}
\EN
where $\beta'=(\beta+\tbeta)/2$ and $\cs^2=\gamma g \HP$ have been used.
We thus find the coefficient $c_0$ as stated in \Eq{urms0}.

\section{Rayleigh number}
\label{Rayleigh}

The purpose of this section is to show that $\epsilon_\ast$,
as defined in \Eq{eps2}, is related to the Rayleigh number $\Ra$.
In laboratory and numerical studies of convection, it is customary
to define $\Ra$ as \citep[e.g.][]{KKB09}
\EQ
\Ra={gd^4\over\nu\chi}
\left(-{\dd\meanS/\cP\over\dd z}\right)_{z_*}^{\rm non-conv}\!\!\!
\approx{gd^4\over\nu\chi}\,\left({\nabrad-\nabad\over \HP}
\right)_{\max}^{\rm non-conv}\!\!\!
\approx{gd^4\over\nu\chi}\,\left({\nabrad-\nabad\over \HP}\right)_{\max},
\label{RaDef}
\EN
which is usually evaluated in the middle of the layer at $z=z_*$.
In the Sun, however, the maximum value is more relevant.
The superscript ``non-conv'' indicates that the entropy gradient
is taken for the non-convecting reference state,
$d$ is the thickness of the layer and $\nu$ is the viscosity.
Introducing the Prandtl number $\Pra=\nu/\chi$ and using the definition
of $\mbox{Nu}$ in \Eq{Nu}, we have
\EQ
{\Pra\,\Ra\over\mbox{Nu}-1}={gd^4\nabad\over\chi^2\HP}.
\EN
On the other hand, using \Eq{eps2}, we find
\EQ
\epsilon_\ast^2={\sigma^3 a_{\rm MLT}\over27\gamma^3}
{(\cs^3/\chi g)^2\over (\kfz\HP)^{4(1-\beta'')}}
={\sigma^3 a_{\rm MLT}\over27 (\kfz\HP)^{4(1-\beta'')}}
{\HP^3 g\over\chi^2}
={\sigma^3 a_{\rm MLT}\,(\HP/d)^4\over27\nabad (\kfz\HP)^{4(1-\beta'')}}
{gd^4\nabad\over\chi^2\HP},
\EN
and therefore
\EQ
\epsilon_\ast^2
={\sigma^3 a_{\rm MLT}/27\nabad \over
(\kfz d)^4 (\kfz\HP)^{-4\beta''} }
\,{\Pra\,\Ra\over\mbox{Nu}-1},
\EN
which shows that $\Ra$ is proportional to $\epsilon_\ast^2$.


\vfill\bigskip\noindent\tiny\begin{verbatim}
$Header: /var/cvs/brandenb/tex/hydro/EntropyRain/paper.tex,v 1.258 2016/10/21 23:57:32 brandenb Exp $
\end{verbatim}


\begin{thebibliography}{}

\bibitem[Arnett et al.(2015)]{Arnett15}
Arnett, W. D., Meakin, C., Viallet, M., Campbell, S. W., Lattanzio, J. C., \& Moc\'ak, M.\yapj{2015}{809}{30}

\bibitem[Asplund et al.(2004)]{Asp04}
Asplund, M., Grevesse, N., Sauval, A. J., Allende Prieto, C.,
\& Kiselman, D.\yana{2004}{417}{751}

\bibitem[Barekat \& Brandenburg(2014)]{BB14}
Barekat, A., \& Brandenburg, A.\yana{2014}{571}{A68}

\bibitem[Basu(1997)]{Basu}
Basu, S.\ymn{1997}{288}{572}

\bibitem[Biermann(1932)]{Bie32}
Biermann, L.\yzfa{1932}{5}{117}

\bibitem[Biermann(1938)]{Bie38}
Biermann, L.\yan{1938}{264}{395}

\bibitem[Blackman \& Field(2003)]{BF03}
Blackman, E. G., \& Field, G. B.\ypf{2003}{15}{L73}

\bibitem[Bogart et al.(2015)]{BBB15}
Bogart, R. S., Baldner, C. S., \& Basu, S.\yapj{2015}{807}{125}

\bibitem[B\"ohm-Vitense(1958)]{Vit58}
B\"ohm-Vitense, E.\yjour{1958}{Z.\ Astrophys.}{46}{108}

\bibitem[Brandenburg et al.(2005)]{BCNS05}
Brandenburg, A., Chan, K. L., Nordlund, \AA., \& Stein, R. F.\yan{2005}{326}{681}

\bibitem[Brandenburg et al.(1996)]{BJNRST96}
Brandenburg, A., Jennings, R. L., Nordlund, \AA., Rieutord, M., Stein, R. F., \& Tuominen, I.\yjfm{1996}{306}{325}

\bibitem[Brandenburg et al.(2004)]{BKM04}
Brandenburg, A., K\"apyl\"a, P. J., \& Mohammed, A.\ypf{2004}{16}{1020}

\bibitem[Brandenburg et al.(2008)]{BRS08}
Brandenburg, A., R\"adler, K.-H., \& Schrinner, M.\yana{2008}{482}{739}

\bibitem[Buske et al.(2007)]{Buske07}
Buske, D., Vilhena, M. T., Moreira, D. M., \& Tirabassi, T.\yjour{2007}{Environ. Fluid Mech.}{7}{43}

\bibitem[Castaing et al.(1989)]{Castaing}
Castaing, B., Gunaratne, G., Heslot, F., Kadanoff, L., Libchaber, A., Thomae, S., Wu, X.-Z., Zaleski, S., \& Zanetti, G.\yjfm{1989}{204}{1}

\bibitem[Cattaneo et al.(1991)]{Cat91}
Cattaneo, F., Brummell, N. H., Toomre, J., Malagoli, A., \& Hurlburt, N. E.\yapj{1991}{370}{282}

\bibitem[Chatterjee \& Antia(2009)]{CA09}
Chatterjee, P., \& Antia, H. M.\yapj{2009}{707}{208}

\bibitem[Christensen-Dalsgaard et al.(1991)]{CDGT91}
Christensen-Dalsgaard, J., Gough, D. O., \& Thompson, M. J.\ysph{1991}{378}{413}

\bibitem[Cossette \& Rast(2016)]{CR16}
Cossette, J.-F., \& Rast, M. P.\yapjl{2016}{829}{L17}

\bibitem[Davidson(2004)]{Dav04}
Davidson, P. A.\ybook{2004}{Turbulence: an introduction for scientists and engineers}{Oxford: Oxford University Press}

\bibitem[Deardorff(1966)]{Dea66}
Deardorff, J. W.\yjas{1966}{23}{503}

\bibitem[Deardorff(1972)]{Dea72}
Deardorff, J. W.\yjgr{1972}{77}{5900}

\bibitem[De Roode et al.(2004)]{DeRoode04}
De Roode, S. R., Jonker, H. J. J., Duynkerke, P. G., \& Stevens, B.\yjour{2004}{Boundary-Layer Meteorology}{112}{179}

\bibitem[Duvall et al.(1997)]{Duvall97}
Duvall, T. L., Jr., Kosovichev, A. G., Scherrer, P. H., Bogart, R. S., Bush, R. I., de Forest, C., Hoeksema, J. T., Schou, J., Saba, J. L. R., Tarbell, T. D., Title, A. M., Wolfson, C. J., \& Milford, P. N.\ysph{1997}{170}{63}

\bibitem[Edwards(1990)]{Edw90}
Edwards, J. M.\ymn{1990}{242}{224}

\bibitem[Ertel(1942)]{Ert42}
Ertel, H.\yjour{1942}{Meteorol. Zeitschr.}{59}{250}

\bibitem[Featherstone \& Hindman(2016)]{FH16}
Featherstone, N. A., \& Hindman, B. W.\yapj{2016}{818}{32}

\bibitem[Freytag et al.(2012)]{Freytag12}
Freytag, B., Steffen, M., Ludwig, H.-G., Wedemeyer-B\"ohm, S., Schaffenberger, W., \& Steiner, O.\yjcp{2012}{231}{919}

\bibitem[Garaud et al.(2010)]{Garaud10}
Garaud, P., Ogilvie, G. I., Miller, N., \& Stellmach, S.\ymn{2010}{407}{2451}

\bibitem[Gizon \& Birch(2005)]{GB05}
Gizon, L., \& Birch, A. C.\yjour{2005}{Liv.\ Rev.\ Sol.\ Phys.}{2}{6}

\bibitem[Gizon \& Birch(2012)]{GB12}
Gizon, L., \& Birch, A. C.\yjour{2012}{Proc. Natl. Acad. Sci.}{109}{11896}

\bibitem[Greer et al.(2015)]{Greer}
Greer, B. J., Hindman, B. W., Featherstone, N. A., \& Toomre, J.\yapjl{2015}{803}{L17}

\bibitem[Grevesse \& Sauval(1998)]{GS98}
Grevesse, N., \& Sauval, A. J.\yjour{1998}{Spa. Sci. Rev.}{85}{161}

\bibitem[Gudiksen et al.(2011)]{Gudiksen}
Gudiksen, B. V., Carlsson, M., Hansteen, V. H., Hayek, W., Leenaarts, J., Mart{\i}nez-Sykora, J.\yana{2011}{531}{A154}

\bibitem[Hanasoge \& Sreenivasan(2014)]{HS14}
Hanasoge, S. M., \& Sreenivasan, K. R.\ysph{2014}{289}{3403}

\bibitem[Hanasoge et al.(2010)]{HBBG10}
Hanasoge, S. M., Duvall, T. L., Jr., \& DeRosa, M. L.\yapjl{2010}{712}{L98}

\bibitem[Hanasoge et al.(2012)]{Hanasoge}
Hanasoge, S. M., Duvall, T. L., \& Sreenivasan, K. R.\ypnas{2012}{109}{11928}

\bibitem[Hanasoge et al.(2016)]{HGS16}
Hanasoge, S., Gizon, L., \& Sreenivasan, K. R.\yanf{2016}{48}{191}

\bibitem[Hathaway et al.(2013)]{Hathaway}
Hathaway, D. H., Upton, L., \& Colegrove, O.\ysci{2013}{342}{1217}

\bibitem[Heslot et al.(1987)]{HCL87}
Heslot, F., Castaing, B., Libchaber, A.\ypra{1987}{36}{5870}

\bibitem[Hill(1894)]{Hil94}
Hill, M. J. M.\yptrsa{1894}{185}{213}

\bibitem[Hotta et al.(2015)]{Hotta}
Hotta, H., Rempel, M., \& Yokoyama, T.\yapj{2015}{803}{42}

\bibitem[Hurlburt et al.(1984)]{HTM84}
Hurlburt, N.E., Toomre, J., Massaguer, J.M.\yapj{1984}{282}{557}

\bibitem[Hurlburt et al.(1986)]{HTM86}
Hurlburt, N. E., Toomre, J., Massaguer, J. M.\yapj{1986}{311}{563}

\bibitem[Hurlburt et al.(1994)]{HTMZ94}
Hurlburt, N.~E., Toomre, J., Massaguer, J.~M., \& Zahn, J.~P.\yapj{1994}{241}{245}

\bibitem[K\"apyl\"a et al.(2009)]{KKB09}
K\"apyl\"a, P. J., Korpi, M. J., \& Brandenburg, A.\yapj{2009}{697}{1153}

\bibitem[K\"apyl\"a et al.(2013)]{KMCWB13}
K\"apyl\"a, P. J., Mantere, M. J., Cole, E., Warnecke, J., \& Brandenburg, A.\yapj{2013}{778}{41}

\bibitem[Kippenhahn \& Weigert(1990)]{KW90}
Kippenhahn, R., \& Weigert, A.\ybook{1990}{Stellar structure and evolution}
{Springer: Berlin}

\bibitem[Lavely \& Ritzwoller(1993)]{LR93}
Lavely, E. M., \& Ritzwoller, M. H.\ysph{1993}{403}{810}

\bibitem[Lord(2014)]{Lord14}
Lord, J. W.\ybook{2014}{Deep convection, magnetism and solar supergranulation}
{PhD thesis, University of Colorado at Boulder}

\bibitem[Lord et al.(2014)]{Lord}
Lord, J. W., Cameron, R. H., Rast, M. P., Rempel, M., \& Roudier, T.\yapj{2014}{793}{24}

\bibitem[Losada et al.(2013)]{Los13}
Losada, I. R., Brandenburg, A., Kleeorin, N., \& Rogachevskii, I.\yana{2013}{556}{A83}

\bibitem[Miesch et al.(2008)]{Miesch08}
Miesch, M. S., Brun, A. S., De Rosa, M. L., \& Toomre, J.\yapj{2008}{673}{557}

\bibitem[Miesch et al.(2012)]{MFRT12}
Miesch, M. S., Featherstone, N. A., Rempel, M., \& Trampedach, R.\yapj{2012}{757}{128}

\bibitem[Mihalas(1978)]{Mih78}
Mihalas, D.\ybook{1978}{Stellar Atmospheres}
{W. H. Freeman: San Francisco}

\bibitem[Moffatt \& Moore(1978)]{MM78}
Moffatt, H. K., \& Moore, D. W.\yjfm{1978}{87}{749}

\bibitem[Nordlund et al.(2009)]{NSA09}
Nordlund, \AA, Stein, R. F., \& Asplund, M.\yjour{2009}{Liv. Rev. Sol. Phys.}{6}{2}

\bibitem[Pleim(2007)]{Ple07}
Pleim, J. E.\yjour{2007}{J. Appl. Met. Climate}{46}{1383}

\bibitem[Priestley \& Swinbank(1947)]{PS47}
Priestley, C. H. B., \& Swinbank, W. C.\yjour{1947}{Proc. Roy. Soc. Lond. A}{189}{543}

\bibitem[Rast(1998)]{Ras98}
Rast, M. P.\yjfm{1998}{369}{125}

\bibitem[Rempel(2004)]{Rem04}
Rempel, M.\yapj{2004}{607}{1046}

\bibitem[Rheinhardt \& Brandenburg(2012)]{RB12}
Rheinhardt, M., \& Brandenburg, A.\yan{2012}{333}{71}

\bibitem[Rieutord \& Zahn(1995)]{RZ95}
Rieutord, M., \& Zahn, J.-P.\yana{1995}{296}{127}

\bibitem[Rieutord et al.(2008)]{Rieutord08}
Rieutord, M., Meunier, N., Roudier, T., Rondi, S., Beigbeder, F., \& Par\`es, L.\yana{2008}{479}{L17}

\bibitem[Rogachevskii \& Kleeorin(2015)]{RK15}
Rogachevskii, I., \& Kleeorin, N.\yjour{2015}{J. Plasma Phys.}{81}{395810504}

\bibitem[Roudier et al.(2012)]{Roudier12}
Roudier, T., Rieutord, M., Malherbe, J. M., Renon, N., Berger, T., Frank, Z., Prat, V., Gizon, L., \& {\v S}vanda, M.\yana{2012}{540}{A88}

\bibitem[R\"udiger(1989)]{R89}
R\"udiger, G.\ybook{1989}{Differential rotation and stellar convection:
Sun and solar-type stars}{Gordon \& Breach, New York}

\bibitem[R\"udiger \& Spahn(1992)]{RS92}
R\"udiger, G., \& Spahn, F.\ysph{1992}{138}{1}

\bibitem[Schou et al.(1998)]{Schou98}
Schou, J., Antia, H. M., Basu, S., Bogart, R. S., et al.\yapj{1998}{505}{390}

\bibitem[Serenelli et al.(2009)]{Ser09}
Serenelli, A. M., Basu, S., Ferguson, J. W., \& Asplund, M.\yapjl{2009}{705}{L123}

\bibitem[Spruit(1974)]{Spr74}
Spruit, H. C.\ysph{1974}{34}{277}

\bibitem[Spruit(1977)]{Spr77}
Spruit, H. C.\yana{1977}{55}{151}

\bibitem[Spruit(1997)]{Spr97}
Spruit, H.\yjour{1997}{Mem. Soc. Astron. Ital.}{68}{397}

\bibitem[Stein et al.(1992)]{SBN92}
Stein, R. F., Brandenburg, A., Nordlund, \AA.\yproc{1992}{148}
{Cool Stars, Stellar Systems, and the Sun}
{M. S. Giampapa \& J. A. Bookbinder}
{ASP Conf. Series, Vol. {\bf 26}}

\bibitem[Stein et al.(2007)]{SBN07}
Stein, R. F., Benson, D., \& Nordlund, \AA.\yproc{2007}{87}
{New Solar Physics with Solar-B Mission, ASP Conf. Ser., Vol. {\bf 369}} 
{K. Shibata, S. Nagata, \& T. Sakurai}
{San Francisco: Astron. Soc. Pac.}

\bibitem[Stein \& Nordlund(1989)]{SN89}
Stein, R. F., \& Nordlund, \AA.\yapjl{1989}{342}{L95}

\bibitem[Stein \& Nordlund(1998)]{SN98}
Stein, R. F., \& Nordlund, \AA.\yapj{1998}{499}{914}

\bibitem[Stein et al.(2009)]{SNGBS09}
Stein, R. F., Nordlund, \AA., Georgoviani, D., Benson, D., \& Schaffenberger, W.\yproc{2009}{421}
{Solar-Stellar Dynamos as Revealed by Helio- and Astero-
seismology: GONG 2008/SOHO 21, ASP Conf. Ser., Vol. {\bf 416}}
{M. Dikpati, et al.}
{San Francisco: Astron. Soc. Pac.}

\bibitem[Stix(1981)]{Sti81}
Stix, M.\yana{1981}{93}{339}

\bibitem[Stix(2002)]{Sti02}
Stix, M.\ybook{2002}{The Sun: An introduction}
{Springer-Verlag, Berlin}

\bibitem[Stull(1984)]{Stu84}
Stull, R. B.\yjas{1984}{41}{3351}

\bibitem[Stull(1993)]{Stu93}
Stull, R. B.\yjour{1993}{Boundary Layer Meteorology}{62}{21}

\bibitem[Trampedach \& Stein(2011)]{TS11}
Trampedach, R., \& Stein, R. F.\yapj{2011}{731}{78}

\bibitem[Tuominen et al.(1994)]{TBMR94}
Tuominen, I., Brandenburg, A., Moss, D., \& Rieutord, M.\yana{1994}{284}{259}

\bibitem[Tuominen \& R\"udiger(1989)]{TR89}
Tuominen, I., \& R\"udiger, G.\yana{1989}{217}{217}

\bibitem[Unno \& Spiegel(1966)]{US66}
Unno, W., \& Spiegel, E. A.\ypasj{1966}{18}{85}

\bibitem[van Ballegooijen(1986)]{vanBall86}
van Ballegooijen, A. A.\yapj{1986}{304}{828}

\bibitem[van Dop \& Verver(2001)]{vDop01}
van Dop, H., \& Verver, G.\yjas{2001}{58}{2240}

\bibitem[Vitense(1953)]{Vit53}
Vitense, E.\yjour{1953}{Z.\ Astrophys.}{32}{135}

\bibitem[V\"ogler et al.(2005)]{VSSCE05}
V\"ogler, A., Shelyag, S., Sch\"ussler, M., Cattaneo, F., Emonet, T., \& Linde, T.\yana{2005}{429}{335}

\bibitem[Woodard(2014)]{Woo14}
Woodard, M.\ysph{2014}{289}{1085}

\bibitem[Woodard(2016)]{Woo16}
Woodard, M.\ymn{2016}{460}{3292}

\bibitem[Xiong \& Deng(2001)]{XD01}
Xiong, D. R., \& Deng, L.\ymn{2001}{327}{1137}

\end{thebibliography}
\end{document}